\documentclass[aps,prd,showkeys,showpacs,amssymb,
amsfonts,epsf,preprintnumbers,nofootinbib,superscriptaddress]{revtex4}

\usepackage[dvips]{graphicx}
\usepackage{bm,latexsym,amsmath,amssymb,amsfonts,color}

\newcommand\beq{\begin{eqnarray}}
\newcommand\eeq{\end{eqnarray}}

\newcommand{\nn}{\nonumber}

\newcommand{\bea}{\begin{eqnarray}}
\newcommand{\eea}{\end{eqnarray}}
\newcommand{\bk}{{\bf k}}
\def\g{{\cal G}}

\begin{document}

\title{Revisiting the spectrum of a scalar field in an anisotropic universe}

\author{Hyeong-Chan Kim}
\email{hckim@ut.ac.kr}
\affiliation{School of Liberal Arts and Sciences, Korea National University of Transportation, Chungju 380-702, Korea}
\author{Masato Minamitsuji}
\affiliation{Yukawa Institute for Theoretical Physics, Kyoto University, Kyoto 606-8502, Japan}

\begin{abstract}
We revisit the issue on signatures of pre-inflationary background anisotropy
by considering the quantization of a massless and minimally coupled scalar
field in an axially symmetric Kasner background, mimicking cosmological perturbations.
We show that the power spectrum of the scalar field fluctuation has a negligible difference from the standard inflation in the non-planar directions, but it has a sharp peak around the symmetry plane.
For the non-planar high-momentum modes, we use the WKB approximation for the first period and the asymptotic approximation based on the de Sitter solution for the next period.
At the boundary, two mode functions have the same accuracy with error of  $O({H_{i}}/k)$.
We calculate the approximation up to the order of $({H_{i}}/k)^6$
and show that the power spectrum of the scalar field fails to get corrections
until we execute the approximation up to $6^{\rm th}$ order.
\end{abstract}
\maketitle

\section{Introduction}

Inflation has become one of the paradigms of modern cosmology.
First of all, inflation elegantly solves many problems which are present in the standard Big-Bang model such as the horizon and
flatness problems.
Second, it accounts for the origin of the large scale structure of the universe in terms of the quantum fluctuations originating
from the adiabatic vacuum structure in early universe.
According to the cosmic no-hair conjecture, the exponential expansion of the universe
during the inflationary period erases
any classical memory~\cite{wald}.
Remarkably, the nature of the primordial fluctuations is understood
in terms of  symmetries of the de Sitter spacetime~\cite{jiro}.
In general, we need $n$-point correlation functions to characterize the statistical nature of primordial fluctuations.
However,  these symmetries lead the power spectrum of a scale invariant form:
\begin{eqnarray} \label{P0}
P_{\phi}^{(0)}=\frac{H_{i}^2}{4\pi^2},
\end{eqnarray}
where ${H_{i}}$ is the Hubble parameter during 
inflation.
These predictions from symmetries are robust and universal in inflationary scenarios.
In fact, the above predictions have been confirmed, e.g.,
 by the measurements of cosmic microwave background (CMB)~\cite{WMAP7}.

As the observational precision increases, we have to go beyond the power spectrum to look at fine structure of the primordial fluctuations.
Since in the realistic inflationary universe the symmetries  of the de Sitter spacetime do not hold exactly, violation of  them provides a measurable effect.
The violation of the temporal de Sitter symmetry leads to a slight tilt of the power spectrum,
which is characterized by a slow roll parameter.
Similarly, the violation of the shift symmetry leads to non-Gaussianities~\cite{non-Gaussianity},
for which we need to prove higher correlation functions.
It is known that observable non-Gaussianities do not appear in the
simplest single field inflaton with a canonical action.
Substantial progresses~\cite{nG:study} have been made in understanding the enhancement in non-Gaussianity for several variants of the standard scenario, involving having multiple scalar fields, non-canonical action for the scalar field, introducing higher derivative terms in the action, or having a non-standard vacuum state.

One may naturally expect violation of the spatial de Sitter symmetry, which would
lead to the statistical inhomogeneity~\cite{inhom:cos}.
Another possibility is introducing spatially homogeneous models violating the spatial isotropy~\cite{ACW,aniso-models,tpu,gkp},
where the Copernican principle is kept since there is no privileged positions in the universe.
Instead, in this model, the universe has a privileged direction.
To reconcile with the spatial homogeneity, the axis should be the same all over the universe.
The existence of a privileged direction which is global over the universe seems unlikely to happen.
It may be possible only when the speciality comes from some quantum mechanical process in the very early universe.
It is known that a primordial anisotropy in the metric will be washed away
during inflation, hence it is interesting to find remaining signatures of the strong anisotropy
in the early spacetime metric in the perturbation spectrum.
From the observational point of view, a lot of anomalies indicating the statistical anisotropy are reported although its statistical significance is uncertain.
Unfortunately, it turns out that many of these models suffer from either instabilities, or a fine tuning problem, or a naturalness problem~\cite{problems}.

We start from the discussion on the evolution of anisotropic universe in the Einstein gravity minimally coupled to a massive scalar field, where the scalar field plays the role of the inflaton.
To obtain a sufficiently long period of inflation, one usually imposes the slow rolling condition,
$\phi_0 \gg M_{P}$, where $\phi_0$ is the initial value of the inflaton.
Under the assumption, as discussed in Ref.~\cite{km2}, the background metric can
be approximated by the Kasner spacetime with a positive cosmological constant, $\Lambda(=3{H_{i}}^2)$.
Among all, we are mainly interested
in the the regular Kasner spacetimes with two dimensional axial symmetry with metric,
\begin{eqnarray} \label{ds2:KdS}
ds^2=-d\tau^2+ \sinh^{\frac23}(3{H_{i}}\tau)
    \left[ \tanh ^{-\frac{2}3}\Big(\frac{3{H_{i}}\tau}2\Big)~ (dx_1^2+ dx_2^2)
    + \tanh ^{\frac{4}{3}}\Big(\frac{3{H_{i}}\tau}2\Big) dx_3^2\right].
\end{eqnarray}
The spacetime has a privileged axis $x_3$ orthogonal to the symmetry plane.
The spacetime has a Rindler-like event horizon at $\tau=0$ since $g_{11}(\,=g_{22})$
approaches to a finite value and $g_{33}$
goes to zero linearly.
The region $\tau\geq 0$ is connected to the region beyond the horizon where the coordinate $\tau$ plays the role of a space coordinate and a timelike singularity is located.
As shown in Ref.~\cite{LKKL}, the geometry is closely connected with the anti-de Sitter black brane solution discovered by Lemos~\cite{Lemos}.
There are various reasons why this spacetime is a good testing ground in analyzing the properties of anisotropic universes.
First, it bears various important features of the whole anisotropic universes including large anisotropy at $\tau=0$.
Second, it is known that this anisotropic inflation is an attractor~\cite{jiro} in supergravity with a wide range of gauge kinetic functions.
This fact will be related to the well-known instability of gravitational wave modes in Kasner universes~\cite{tpu,gkp}, since the instability
becomes controllable only for the above case~(\ref{ds2:KdS}).
Third, only in this symmetric case of all anisotropic expansions, we can impose proper anisotropic vacuum state in terms of
the zeroth order WKB approximation~\cite{km}.

If one uses the Sasaki-Mukhanov variable, except for the complication
due to the mixing of tensor-scalar modes, the evolutions of the metric perturbations
are not much different from that of a massless scalar field~\cite{gcp,Paban}.
In Ref.~\cite{km2}, it was also shown that the mode mixing modifies the power spectrum only
by a proportionality factor with a small correction term of 
order $m/{H_{i}}$, where $m$ is the mass of the inflaton.
Therefore, as a formulation level, it is
good to deal with the scalar field
rather than the metric perturbation itself.
Hence in this work, we are interested in the evolution of a massless,
minimally coupled scalar field propagating on the background anisotropic universe~(\ref{ds2:KdS}) with action
\beq \label{S:phi}
S_{\phi}=-\frac{1}{2}
\int d^4 x\sqrt{-g}
\Big(g^{\mu\nu}\partial_{\mu}\phi\partial_{\nu}\phi
\Big).
\eeq
This scalar field is not the inflaton mentioned above but just a {\it mimic of 
metric perturbations}.
Hence it is assumed that its size is small enough so that it does not modify the background dynamics.
We focus on the quantization procedure and the evolution in the anisotropic stage.
The effects of a scalar field onto the geometry of the anisotropic Universe were discussed in \cite{ms}.
The canonical quantization of the scalar field can be done in the standard manner.
The vacuum is chosen at the initial anisotropic era: $\tau \to +0$.
Usually, in a time-dependent quantum harmonic oscillator, an adiabatic process implies the one where the potential changes slowly enough compared to its size, and the time evolution can be obtained from the WKB approximation.
In the standard inflationary models, an adiabatic vacuum in the early universe is also defined in the same way.
However, in anisotropic space-times, the situation gets worse for two reasons.
First, most modes do not take oscillatory form, therefore, the adiabatic vacuum does not exist.
Second, there exists the instability problem for the tensor modes of gravitational perturbations.
Only in the specific solution~(\ref{ds2:KdS}), the two problems disappear and the adiabatic vacuum exists, which was called an {\it anisotropic vacuum} in Ref.~\cite{km}.
It is important to note that, in this scheme, the entire information of anisotropy (and therefore, the cosmological structure at the present universe) is encoded in the vacuum state of the theory
similarly to the case of the Robertson-Walker space-time.

In general, the evolution of the scalar field in this spacetime cannot be solved exactly.
To have an analytic form of the spectrum, we develop an appropriate approximation method.
Usually, the WKB approximation and the de Sitter solutions are commonly used at the early and later stages
of evolutions~\cite{km,km2,Paban}.
For ordinary high momentum modes with $k\gg {H_{i}}$, the WKB approximation is valid.
However, at the very early stage of the anisotropic expansion,
the validity condition for the WKB approximation changes to $k_3\gg {H_{i}}$,
where $k_3$ is the comoving momentum along the special axis.
Therefore, there exist non-trivial high momentum modes (the planar modes, $k_3 \sim {H_{i}}$ and $k\gg {H_{i}}$)
which cannot be dealt with the WKB approximation but need separate treatments.

Does the present approximation successfully predict the power spectrum for the present universe?
For the observed wave length scales
smaller than $10^4 \mbox{Mpc}$, the mode with comoving momentum $k_{\rm obs}$
can be seen if it satisfies~\cite{km,Paban}
\beq
\label{limit}
  k_{\rm obs}> k_{\rm crit} \equiv
    {H_{i}} \, e^{N-64}\left(\frac{T_R}{10^{14}\mbox{GeV}}\right)
        \left(\frac{10^{16} \mbox{GeV}}{V_{\rm inf}^{1/4}}\right)^2 ,
\eeq
where $k_{\rm crit}$, $N$, $T_R$, and $V_{\rm inf}$
are the critical observable comoving wavenumber, the number of $e$-folding,
the reheating temperature and the value of inflaton potential, respectively.
Thus, for inflation in the grand unified theory (GUT)
if $e$-folding is around $N \gtrsim 64$, the modes may satisfy $k_{\rm obs}\gtrsim H_{i}$.
Therefore, most observable modes in our universe marginally satisfy the WKB approximation condition.
In this sense, the present approximation can successfully expect the evolutions of all non-planar modes.
For a sufficient long duration of inflation, say $N \gg 64$, all observable modes
satisfy $k\gg H_{i}$
and the WKB approximation holds well.
However, in this case, it will be very hard to find any signature originating from the initial anisotropy since most correction terms will be suppressed
by some power of $H_{i}/k \sim e^{64-N}$ except for the planar modes.
If $N\simeq 60$, the modes which belong to the wavelength scale between $10^2\,\mbox{Mpc}$ and $10^4\,\mbox{Mpc}$, will 
bear much information about the anisotropic preinflationary memory, which must be already detected.
This may constrain the $e$-folding of the inflationary expansion from being smaller than $64$.
Assuming $N\geq 64$, it is not difficult to satisfy the WKB condition for the range of observable scales.
However, the visibility of the anisotropy at these scales is not guaranteed.

In this work, we show that the relic of the preinflationary anisotropy appears to be very difficult to be observed in the non-planar modes except
for the special case that the residue of the comoving momentum satisfying $H_i< k<10 H_i$ are located in the observable range.
This is simply because the non-vanishing correction to the power spectrum comes from very high order perturbative corrections and therefore, its size is suppressed by high power of $H_i/k$.
Perturbative calculations on the power spectrum were done previously in Ref.~\cite{km,km2,Paban}, where the authors are missing doing perturbative approximation for the asymptotic solution around the de Sitter one.
Still, there are possibilities that the anisotropic effect can be detected
only from the power spectrum for planar modes.

In Sec. II, we quantize the scalar field and gives the initial condition for the scalar evolution by setting the mode solution to be purely positive frequency at early times.
In Sec. III, we introduce the approximation method used in this work and show the resulting power spectrum of the scalar field up to $6^{\rm th}$ order of the expansion parameter.
We show that the non-trivial correction to the power spectrum from initial anisotropy survive only from that order.
In Sec. IV, we present the power spectrum for the case of a planar modes where the contribution from the initial anisotropic expansion of the background geometry appears in its zeroth order approximation.
In Sec. V, we summarize the results.
We attach a few Appendices where the details of the calculations are given.

\section{Quantization of the Scalar field on anisotropic vacuum}
The canonical quantization of the minimally coupled massless scalar field
with the action~(\ref{S:phi}) in the anisotropic spacetime~(\ref{ds2:KdS}) is done
in the standard manner:
\beq
\label{mode_dec}
\phi= \int d^3k
    \Big(u_{\bf k} a_{\bf k}
+    u^{\ast}_{\bf k} a^{\dagger}_{\bf k}
\Big),
\eeq
where the creation and annihilation operators satisfy
the commutation relations $\big[ a_{\bf k_1},a^{\dagger}_{\bf k_2}\big]
=\delta({\bf k_1}-{\bf k_2})$ (others are zero)
and $u_{\bf k}= e^{i{\bf k}{\bf x}}\phi_{\bf k}/(2\pi)^{3/2}$.
The details of the quantization process depend on the choice of the mode function $u_{\bf k}$.
We normalize the mode function as
\begin{eqnarray}\label{normalization}
\phi_{\bf k}\partial_\tau \phi^{\ast}_{\bf k}
-\big(\partial_\tau \phi_{\bf k} \big)\phi^{\ast}_{\bf k}
=\frac{i}{e^{3\alpha}}.
\end{eqnarray}

For the later convenience, we introduce a dimensionless time $x$ by
\begin{eqnarray} \label{x:tau}
\sinh(\varepsilon x)= \frac{1}{\sinh (3 {H_{i}}\tau) }=e^{-3\alpha}\,,
\end{eqnarray}
where $\varepsilon$ denotes a small expansion parameter which will be specified later.
The arrow of time for $x$ is inverted since it varies from $\infty$ to $0+$
as the comoving time $\tau$ increases from $0+$ to $\infty$.
The asymptotic limit with $\tau \gg {H_{i}}^{-1}$ is now equivalent to $\varepsilon x\ll 1$.
Therefore, the asymptotic expansion in the long time limit corresponds to the $\epsilon$ expansion.

The equation of motion for the scalar field is written in the form of a time-dependent oscillator
\beq \label{eom}
\Big(\frac{d^2}{dx^2}+\Omega_{\bf k}(x)^2 \Big)\phi_{\bf k} =0\,,
\eeq
where the dimensionless frequency squared is
\beq
\Omega_{\bf k}^2(x)
=\left(\frac{\varepsilon}{3{H_{i}}}\right)^2\frac{2^{4/3}\big(k_\perp^2
    e^{-2\varepsilon
    x}+k_3^2\big)}  {(1-e^{-2\varepsilon x})^{4/3}}
=\frac{2(\bar k\varepsilon^{2/3})^2}{9}\,\left(\frac{e^{\varepsilon x}}{\sinh\varepsilon x}\right)^{1/3} \left(\frac{1}{e^{2\varepsilon x}-1}+r^2\right)  \,.
\eeq
Here we define a scaled wave-number and a measure of planarity of a given mode by
\begin{eqnarray}
\bar k =\varepsilon^{1/3}\frac{ k}{{H_{i}}}\,, \qquad r=\frac{k_3}{k}\,,
\end{eqnarray}
where $k^2:= k_1^2+k_2^2+k_3^2=k_\perp^2+k_3^2$. Later in this paper, we omit ${\bf k}$ in the frequency squared $\Omega_{\bf k}^2$ for simplicity.
In terms of $x$, the normalization condition~(\ref{normalization}) changes to
\begin{eqnarray}\label{wronskian:x}
\phi_{\bf k} \partial_x \phi_{\bf k}^*-(\partial_x \phi_{\bf k})\phi_{\bf k}^* =-\frac{i\varepsilon}{3{H_{i}}}.
\end{eqnarray}
 Here, we use $d\tau/dx= - e^{3\alpha}\times\varepsilon/(3{H_{i}})$.

The power spectrum is defined by
\beq
\big\langle 0| \phi^2 |0\big \rangle
:=\int d\ln k\, \int\frac{d\theta_{\bf k}}{2} P_{\phi},\quad
P_{\phi}=\frac{k^3}{2\pi^2} \big|\phi_{\bf k} \big|^2  .
\label{ps}
\eeq
In contrast to the case of the standard 
de Sitter universe
the direction dependence would be included in the power spectrum.
The vacuum is chosen at the initial anisotropic era: $\tau \to +0$ to satisfy $a_{\bf k} |0\rangle =0$.
For this purpose, we choose the solution to be purely positive frequency mode with respect to $\tau$ at the early stage:
\begin{eqnarray} \label{phi:initial}
\lim_{\tau \to 0}\phi_{\bf k} = \frac{1}{\sqrt{2\times 2^{2/3} k_3}} e^{i\omega x+i\psi},
\end{eqnarray}
where we follow the normalization~(\ref{normalization}) and $\omega = \frac{2^{2/3}}{3}\frac{\varepsilon k_3}{{H_{i}}}$.

\section{
Non-planar high-momentum modes
} \label{sec:approx}
As mentioned in the introduction, we use the WKB solution at early times ($x>x_\ast$)
and
asymptotic solution at later times ($x<x_\ast$).
At the matching time $x=x_\ast$, the accuracies of the two are equal.
Later in this paper, we set $x_\ast =1$ by choosing $\varepsilon$ appropriately.
We also assume that $\bar k$ is larger than one, which will be satisfied with the modes we are interested in at the present approximation.

\subsection{WKB solution at the early times}
The WKB approximation is valid up to the order of the correction term if
\begin{eqnarray}\label{WKBcondition}
\epsilon(x):= \left|\frac{\frac{d\Omega^2(x)}{dx}}{\Omega^3(x)}\right|
= \frac{1}{\bar k}
   \left(\frac{2\varepsilon}{1-e^{-2\varepsilon x}}\right)^{1/3}
   \frac{e^{-2\varepsilon x}}{\sqrt{1+e^{-2\varepsilon x}}}
   \frac{1 + \frac{1+r^2}2 \tanh \varepsilon x}{
    \left[\frac{1}{e^{2 \varepsilon x}+1} +r^2 \tanh \varepsilon x\right]^{3/2} } \ll 1 ,
\end{eqnarray}
where $\epsilon$ plays the role of the adiabatic parameter.
For most modes, the adiabaticity condition is valid if $\varepsilon x \gg 1$ because of the exponentially decaying factor $e^{-2\varepsilon x}$.
For a small $\varepsilon x$, the condition is also valid for non-planar high momentum modes if $x \gg \bar k^{-3}$.
In the asymptotic regime with $0\leq x<\bar k^{-3} $, the adiabaticity condition fails to be satisfied.
On the other hand, the asymptotic (small $\varepsilon x$) approximation is valid if $ x \ll \varepsilon^{-1} $.
The asymptotic expansion in this case will be dealt in the next subsection.

The WKB solution satisfying the initial condition~(\ref{phi:initial}) is given by
\beq
\phi_{\rm WKB}=\sqrt{\frac{\varepsilon}{3{H_{i}}}}\,\frac{1}{\sqrt{2\tilde \Omega(x)}}
\left\{\exp\Big[i\int_{x_0}^x dx'\tilde \Omega (x') +i \psi\Big] \right\},
\eeq
where we choose the positive (negative) frequency mode with respect to $t$ ($x$) and $\psi$ is a phase factor. 
$\tilde \Omega$ satisfies the nonlinear equation
\begin{eqnarray} \label{WKB:iteration}
\tilde \Omega (x)^2=\Omega (x)^2\,c(x)^2; \qquad c(x)^2 :=1
   -\frac{\tilde \Omega_{,xx}}{2\Omega^2\tilde \Omega}
        +\frac34\left(\frac{\tilde \Omega_{,x}}{\Omega(x)\tilde \Omega(x)}\right)^2.
\end{eqnarray}
The WKB wavefunction is expanded up to an enough adiabatic order,
to validate our matching scheme with the solutions in the asymptotic region.
Noting Eq.~(\ref{WKB:iteration}), the next order of the approximation will improve the accuracy by the order 
\begin{eqnarray} \label{E:WKB}
E_{\rm WKB}(x)\sim \epsilon(x)^2 \simeq \frac{1}{\bar k^2 \, x^{2/3}}.
\end{eqnarray}
For example, to the order $\bar k^{-4}$, we get
$
c(x)^2= 1-\frac{2}{\bar k^2 x^{2/3}} +\frac{3}{\bar k^4 x^{4/3}}
    +\frac{3r^2-1}{3} \frac{\varepsilon x^{1/3}}{\bar k^2}+\cdots,
$
where the last term has the mixed order of magnitude $\varepsilon/\bar k^2$.

\subsection{Later time solution}

In the later time limit, the space-time approaches that of the de Sitter spacetime.
Therefore, it is natural that the zeroth order solution is just the well-known scalar field mode solution in the de Sitter spacetime.
The higher order corrections are based on the {\it asymptotic approximation} based on the limit $\varepsilon x \ll 1$.
By series expanding the frequency squared, we approximate the equation of motion~(\ref{eom}) to be
\begin{eqnarray} \label{eom:series}
\left(\frac{d^2}{dx^2}+\sum_{n=0}^\infty\varepsilon^n  V_n \right)\phi=0\,,
\end{eqnarray}
where the order by order corrections of the frequency squared are
\begin{eqnarray}
&&V_0=\frac{\bar k^2}{9x^{4/3}},\qquad
    V_1=-\frac{2\bar k^2 q^2}{9 x^{1/3} }, \qquad
    V_2=\frac{2\bar k^2 r^2 x^{2/3}}{27} , \qquad
    V_3 =\frac{8 \bar k^2 x^{5/3}}{729}, \cdots
\end{eqnarray}
with $q^2=1/3 - r^2$. For higher orders than these, refer Eq.~(\ref{Vi}) in Appendix~\ref{app:B}.
Note that each $V_i$ is of the same order as $V_0$ in this expansion. This is why we prefer $x$ to $\tau$.
We present the corrections up to the $6^{\rm th}$ order in $\varepsilon$ in Appendix~\ref{app:B}.
This is because nontrivial corrections to the power spectrum of the isotropic case
exist from this order.

The general solution to the differential equation~(\ref{eom:series}) is given by 
\begin{eqnarray}
\phi = A_+ u(x) + A_-v(x)\,,
\end{eqnarray}
where $u$ and $v$ are the positive and negative frequency modes, respectively.
We get the approximate solution by series expanding the modes $u=u_0 + \varepsilon u_1 +\varepsilon ^2 u_2+ \cdots$ and $v=v_0+\varepsilon v_1+\varepsilon^2 v_2+\cdots $ and then solving the equation of motions order by order.
The zeroth order equation satisfies
\begin{eqnarray}
&&\left(\frac{d^2}{dx^2} +\frac{\bar k^2}{9x^{4/3}}\right)\phi_0=0.
\end{eqnarray}
Its solutions are nothing but the de Sitter mode solutions
\begin{eqnarray} \label{uv}
u_0(x):=\Big(-1+i \bar k x^{1/3}\Big)
e^{i \bar k x^{1/3}}, \quad v_0(x):=\Big(-1-i \bar k x^{1/3}\Big)
e^{-i \bar k x^{1/3}}=u_0^*(x),
\end{eqnarray}
where the Wronskian becomes $u_0(x) v_0'(x) -u_0'(x) v_0(x)  = -2i\bar k^3/3$.
This Wronskian condition will be kept even if we get higher order corrections.

The equation of motion for the first order correction term is
\begin{eqnarray}
&&\left(\frac{d^2}{dx^2} +\frac{\bar k^2}{9x^{4/3}}\right)\phi_1 =
  -V_1(x) \phi_0(x) .
\end{eqnarray}
The left hand side of the equation of motion is the same as that of the zeroth order
and $-V_1(x)\phi_0(x)$ 
plays the role of a source.
Its solution is formally given by the integral,
\begin{eqnarray}
\phi_1(x) &=&-\int_0^{x_c} G(x,y) V_1(y)\phi_0(y) dy,
\end{eqnarray}
where the Green's function is,
\begin{eqnarray} \label{greenfn}
G(x,y) = -\frac{3}{2i\bar k^3} \left[u_0(x)v_0(y) \theta(y-x) +u_0(y)v_0(x)\theta(x-y)\right].
\end{eqnarray}
One may choose the upper bound of the integration $x_c$ to be any 
value greater than $x$ formally.
In fact, the choice of $x_c$ changes only the homogeneous part of the solution (explicitly $\phi_0$), which is not relevant in doing the perturbative higher order calculation.
Therefore, we simply drop all $x_c$ dependent term after the integration so that the first order corrections
$u_1$ 
is the complex conjugate to $v_1$.
Inserting $\phi_0(y) \to u_0(y)$ or $\to v_0(y)$ we get
\begin{eqnarray}
v_1^*(x)=u_1 (x) &=&
\frac{q^2 e^{i \bar k \sqrt[3]{x}} \left(2 \bar k^5 x^{5/3}+6 i \bar k^4 x^{4/3}-6 \bar k^3 x-9 \bar k \sqrt[3]{x}-9 i\right)}{8 \bar k^3}-\frac{9 q^2 e^{-i\bar k
   \sqrt[3]{x}} \left(\bar k \sqrt[3]{x}-i\right)}{8 \bar k^3}\,.
\end{eqnarray}
One may easily check that the first order solutions $u\simeq u_0+ \varepsilon u_1$ and $v\simeq v_0+\varepsilon v_1$ satisfy the Wronskian condition up to the order of the calculation.
Note that $u_1(x)$ vanishes at $x=0$, which happens because we dropped all $x_c$ dependent terms.

Now, let us check how much the accuracy 
is improved if we use the first order solutions, compared to the zeroth order one.
The dominant term in $u_1$ in the high $\bar k$ limit is $O(\bar k^2)$.
The first order correction term is $O(\varepsilon \bar k^2)$ and the zeroth order term is $O(\bar k)$.
Therefore, the accuracy of the solution is increased by $\varepsilon \bar k$, which should be a small number for the approximation to be valid.
To determine the time when the WKB and the asymptotic approximations will be matched, we need to know the size of the error of the zeroth order solution for a given $x$.
The relative ratio of the correction term to the zeroth order solution gives the error,
\begin{eqnarray} \label{E:asym}
E_{\rm asym}(x) \sim \varepsilon \bar k \, x^{4/3} .
\end{eqnarray}
For high frequency modes, there is a long period where both approximations are applicable:
\begin{eqnarray} \label{cond:x*}
\bar k^{-3}\ll x \ll \varepsilon^{-1}.
\end{eqnarray}
To minimize the error of the approximation, we choose the time $x_\ast$ and use the WKB wavefunction for $x>x_\ast$
and the asymptotic solution for $0<x<x_\ast$.
The approximation will be best if we choose the intermediate time $x_\ast$
so that the accuracies of the two solution match:
$$
E_{\rm WKB}=E_{\rm asym},
$$
which gives $x_\ast= (\varepsilon \bar k^3)^{-1}$. For simplicity, we choose to set $x_\ast =1$. Therefore, the small expansion parameter becomes
\begin{eqnarray} \label{t_ast}
\varepsilon= \bar k^{-3}=\left(\frac{{H_{i}}}{k}\right)^{3/2}.
\end{eqnarray}
Now, the size of error at $x=x_\ast$ becomes $E_{\rm asym} = \bar k^{-2}={H_{i}}/k$, which ensures that the present approximation works well for high momentum modes.

In this subsection, we illustrate the calculation up to $O({H_{i}}/k)^3$.
To do this, we need to get the perturbative solution up to the third adiabatic order in $\varepsilon$. The second and third order corrections are given by the integrals,
\begin{eqnarray}
u_2(x) &=& -\int_0^{x_c} G(x,y)[u_0(y)V_2(y)+ u_1(y)V_1(y)] dy, \nn \\
u_3(x) &=& -\int_0^{x_c} G(x,y)[u_0(y)V_3(y)+ u_1(y)V_2(y)+u_2(y) V_1(y)] dy.
\end{eqnarray}

Therefore, up to $O(\bar k^{0})$, we get the second order corrections
$$
u_2(x) = \frac{\bar k^3x^3 e^{i \bar k x^{1/3}}}{32}\left[-i q^4+
    \frac{51 q^4+\frac{32 q^2}{3}-\frac{32}{9}}{7 \bar k \sqrt[3]{x}}+ \left(\frac{99 q^4}{8}+3 q^2-1\right)\left(\frac{16 i}{7 \bar k^2 x^{2/3}}
    -\frac{32 }{5 \bar k^3 x}\right)+
     \cdots\right].
$$
The third order corrections up to $O(\bar k^3)$ 
are given by the integral,
$$
u_3(x) =
-\frac{\bar k^4q^2}{384} e^{i \bar k x^{1/3}} x^{13/3}\left[q^4+\frac{i \left(291 q^4+96 q^2-32\right)}{21 \bar k
   \sqrt[3]{x}}+O\left(\frac{1}{\bar k^2}\right)\right],
$$
In summary, the solution to order $O(\bar k^{-6})$ becomes
\begin{eqnarray}
u(x) &\simeq &u_0(x) +\varepsilon u_1(x) +\varepsilon^2 u_2(x) + \varepsilon^3 u_3(x).
\end{eqnarray}
We have obtained the perturbative solution up to 6$^{\rm th}$ order in $\varepsilon$ in Appendix~\ref{app:B},
where the nontrivial direction dependent correction to the power spectrum appears for the first time.

\subsection{Matching} \label{sec:3C}
The WKB solution is now matched to the asymptotic solution in the asymptotic region, at $x=1$.
We illustrate the calculation to the order of $({H_{i}}/k)^3$.
Calculation to higher order to $({H_{i}}/k)^6$ is given in Appendix~\ref{app:C}.
Remember that the accuracies of the two solutions become equal at $x=1$.
Therefore, on the whole, the present approximation becomes the order by order expansion with respect to $\varepsilon \bar k=\bar k^{-2}= H_i/k$.
The matching condition is given by
\beq
&&
A_+ u(1)+  A_- v(1)=\phi_{\rm WKB}(1) =
  \sqrt{\frac{\varepsilon}{3{H_{i}}}}\frac{\Phi_{\ast}}{\sqrt{2\tilde\Omega_{\ast}}} ,
\nonumber\\
&&
A_+ u{}'(1)+  A_- v{}'(1)=\phi_{\rm WKB}'(1) =i
    \left( 1+\frac{ i\tilde\Omega(x_\ast)_{,x}}{ 2\tilde \Omega_\ast^2}   \right) \sqrt{\frac{\varepsilon}{3{H_{i}}}}\sqrt{\frac{\tilde \Omega_\ast}{2}}\Phi_{\ast}\,,
\eeq
where the phase factor is given by
\beq
\Phi_{\ast}
={\rm exp}
\Big[
i \int_{x_0}^{1}\tilde \Omega dx+i \psi
\Big].
\eeq
The frequency at $x_\ast$ is given by $\Omega_* = \frac{\bar k}{3}\left[1-q^2 \varepsilon+\frac{1-12 r^2 +9r^4 }{18}\varepsilon^2 +O(\varepsilon^3)\right]$ and the ratio of the WKB frequency with respect to $\Omega_\ast$ and its first derivative are
\beq
c(x_\ast)&=&1-\frac1{\bar k^2}+\frac1{\bar k^4}+\frac{3r^2-1}{\bar k^5}-\frac{1}{\bar k^6}
+O(\frac{1}{\bar k^{7}})\,, \nn \\
\frac{\tilde \Omega(x_\ast)_{,x}}{\tilde \Omega^2_\ast} &=&-\frac{2}{\bar k}\left(1+\frac{5q^2}{2\bar k^3}+\frac{9 q^2}{4 \bar k^5} +O(\bar k^{-6}) \right) \,.
\eeq

From the Wronskian relations~(\ref{wronskian:x}) and (\ref{wronskian:uv}), it is straightforward to confirm that the coefficients $A_+$ and $A_-$ satisfy the normalization condition, to the present order,
\beq
\big|A_{+}\big|^2-\big|A_{-}\big|^2
=\frac{\phi \partial_x\phi^*-(\partial_x\phi)\phi^*}{uv'-u'v}=
    \frac{\varepsilon}{2{H_{i}}\, \bar k^3}\,.
\eeq
As the reference,
in the standard isotropic 
initial universe, the normalized mode function is given by
\beq
\phi_{\rm isotropic}
=\frac{i {H_{i}}}{\sqrt{2k^3}}(-1-i\bar k x^{1/3})e^{i\bar k x^{1/3}} .\label{iso}
\eeq

\subsection{Power spectrum}

In the limit of the later times $x\to 0$, using $u_{i}(0)=0$ with $i=1,2,3\cdots$, the field behaves as
\beq
 \phi_{\rm fin}=\phi\Big|_{x\to 0}
&=&-(A_+ + A_-)= \frac{3i}{2\bar k^3}\left[\phi_{\rm WKB}(1)(u'(1)-v'(1))-
     \phi_{\rm WKB}'(1)(u(1)-v(1))\right]\nn \\
&=& i \sqrt{\frac{3\varepsilon}{2{H_{i}} \bar k^3}} \frac{\Phi_\ast}{\sqrt{2}}
    \left\{\sqrt{\frac{\bar k}{\tilde \Omega_*}}\frac{u'(1) - v'(1)}{\bar k^2}-
        i\left(1+\frac{i\tilde\Omega'(x_\ast)}{ 2\tilde\Omega_*^2} \right) \sqrt{\frac{\tilde\Omega}{\bar k}}\, \frac{ u(1)-v(1)}{\bar k} \right\}
. \nn
\eeq
The explicit value to $O(\bar k^{-6})$ becomes
\begin{eqnarray} \label{phi:fin}
\phi_{\rm fin} &=&\sqrt{\frac{\varepsilon}{2{H_{i}}\bar k^{3}}}
 \Phi_\ast e^{-i \bar k}
    \left[ i+\frac{1}{\bar k}-\frac{\frac{q^2}{2}+i}{2\bar k^2}
        +i\frac{\frac{ q^2}{2}+i}{2\bar k^3}
    -\frac{i \left(q^4+12 i q^2-12\right)}{32 \bar k^4}+\frac{-207 q^4-(96+756 i)
   q^2+788}{2016 \bar k^5}\right. \nn \\
&&\left.+\frac{21 q^6+1206 i q^4-(3780-384 i) q^2-2648 i}
    {8064 \bar k^6}+O\left(\frac{1}{\bar k^7}\right)\right].
\end{eqnarray}
Note that the final form of the mode solution gains direction dependent corrections order by order in $\bar k$.
However, the power spectrum~(\ref{ps}) obtained from the field~(\ref{phi:fin}) fails to gain corrections from the standard de Sitter spectrum to the present order:
\beq
P=\frac{{H_{i}}^2}{4\pi^2} +O\Big(\Big(\frac{{H_{i}}}k\Big)^4\Big).\nn
\eeq

As shown in the calculation in Appendix~\ref{app:C}, the power spectrum acquires corrections only when we calculate to the adiabatic order $O(\bar k^{-12})$.
The power spectrum including the corrections becomes
\begin{eqnarray} \label{P:phi}
P
&=& \frac{{H_{i}}^2}{4\pi^2}
  \left\{
  1 + \frac{9(11-90r^2+99 r^4)}{32}\left(\frac{{H_{i}}}{k}\right)^6 +O\left(\Big(\frac{{H_{i}}}{k}\Big)^{7}\right) \right\} \,.
\end{eqnarray}
For mode with $k\sim 10 {H_{i}}$, the relative size of the correction term is of $O(10^{-6})$.
The isotropy violation at initial stage of the universe is not small but its effects
on the spectrum for the non-planar modes are suppressed
by the long duration of inflation and high momentum effect.
The correction $
O( (H_i/k)^6)$ is highly dependent on $k$ to suppress the anisotropy effect.

Before closing this section, it is interesting to compare the power spectrum obtained above with the parametrization done by Ackerman, Carroll and Wise~\cite{ACW}.
They expected that the power spectrum in a generic model of inflation with broken rotational invariance  will be given by
$$
P(k)=
P_0(k)
\left\{1+ g(k) \cos^2\theta
\right\},
$$
where $P_0(k)$ is the usual scale-invariant contribution, $r=\cos\theta=k_3/k$,
and $\theta$ denotes the angle between the preferred direction
and the wavenumber vector of fluctuations.
Our form of the power spectrum is a generalized form of it. 
There are corrections to the direction independent term and they include higher order terms in $\cos\theta$, i.e., $\cos^4\theta$ term, not only the quadratic component $\cos^2\theta$.
Thus higher order terms in $ \cos\theta$ are not suppressed.

\section{Planar modes}

\subsection{Power spectrum}
For the planar modes $r^2\sim 0$, there appears a region where the WKB approximation may not be valid during a period in $\varepsilon x \gg 1 $.
We divide the time into three separate regions divided by the times $x_{1}$ and $x_{\ast}$.
In the region $x_1>x> x_{\ast}$ the WKB approximation is valid.
For other two regions, we may find approximate solutions.
In the case of $r=0$ exactly, the adiabaticity parameter diverges
in the limit of $x\to \infty$ and there is no anisotropic vacuum state.
The mode $r=0$ would behave classically, not quantum mechanically,
and will be out of scope in this paper.

\begin{figure}[tbph]
\begin{center}
\begin{tabular}{ll}
\includegraphics[width=.5\linewidth,origin=tl]{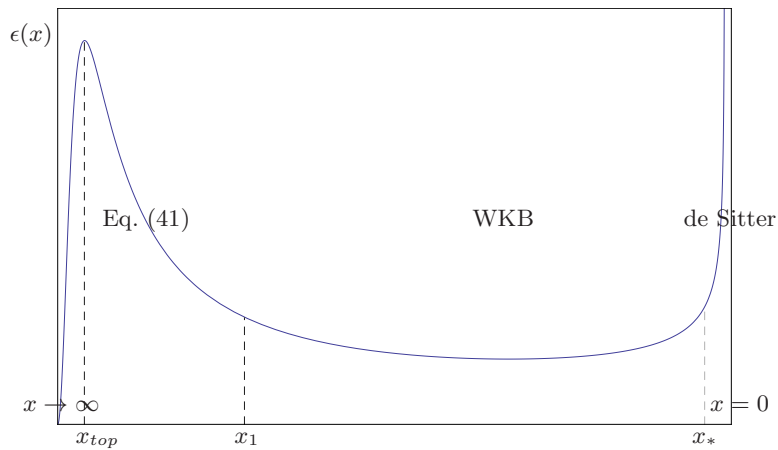}
\end{tabular}
\put (-275,70) {$\epsilon(x)$}
\put (-20,-82) { $x_\ast$}
\put (-190,-82) {$x_1$}
\put(-250,-82) {$x_{top}$}
\put(-270,-70) {$x\to \infty$}
\put(-10,-70)  {$x = 0$}
\put(-240,0) {Eq.~\eqref{phi11}}
\put(-100,0){WKB}
\put(-20,0) {de Sitter}
\end{center}
\caption{ The behavior of the adiabatic parameter in the case of a planar mode.
The left hand side ($x\to \infty$) corresponds to the past.
The WKB approximation is temporary violated around $x\simeq x_{top}$.
} \label{fig:p4}
\end{figure}

The WKB approximation is valid up to the order of the correction term if
$$
\epsilon(x):= \left|\frac{\frac{d\Omega^2(x)}{dx}}{2\Omega^3(x)}\right|=
    \frac{1}{\bar k}\left(\frac{2\varepsilon}{1-e^{-2\varepsilon x}}\right)^{1/3}
    \frac{e^{-2\varepsilon x}}{\sqrt{1+e^{-2\varepsilon x}}}
    \frac{1+\frac{1+r^2}{2} \tanh\varepsilon x}{\left[\frac{1}{e^{2\varepsilon x}+1}+r^2\tanh\varepsilon x\right]^{3/2}} \ll 1.
$$
The characteristic behavior of $\epsilon(x)$ is shown in Fig.~\ref{fig:p4}.

We are considering the planar mode with $r\ll 1$.
For a small $r \ll e^{-\varepsilon x}$, the adiabatic parameter becomes of
$$
\epsilon(x)\simeq \frac{H_i}{ k \,e^{-\varepsilon x}}\frac{3+e^{-2\varepsilon x}}{2^{2/3}(1-e^{-2\varepsilon x})^{1/3}} +O(r^2)
$$
We notice, for a large $x$, that the value of $\epsilon(x)$ should satisfy
\begin{eqnarray} \label{x1:kh}
\epsilon(x)\simeq \frac{3H_i}{2^{2/3}k e^{-\varepsilon x}} \ll 1 \longrightarrow e^{\varepsilon x_1} \ll \frac{k}{H_i}.
\end{eqnarray}
Therefore, $e^{\varepsilon x_1}$ must be large but should be smaller than $k/H_i$ if the WKB approximation holds.
In addition, we may also expand $\epsilon(x)$ for a large $x$:
$$
\epsilon(x\gg 1) \simeq \frac{3 \varepsilon^{1/3} e^{-2\varepsilon x}}{2^{2/3}\bar k r^3}
    \left(1+\frac{r^2}{3} -\frac{27-30r^2-5r^4}{18}\frac{
e^{-2\varepsilon x}}{r^2}+\cdots \right).
$$
For a small $e^{-\varepsilon x}$, according to this expansion, the adiabatic parameter increases as $e^{-2\varepsilon x}$ until $e^{-\varepsilon x}< r\ll 1$.
It has a peak at $x=x_{top}$ with $e^{-\varepsilon x_{top}}\sim r$ when when the $O(e^{-4\varepsilon x})$ term becomes equally important with the dominant term.
Later it starts to decrease.
Therefore, we additionally need 
\begin{eqnarray} \label{eex:r}
e^{\varepsilon x_1} \ll \frac 1r\,,
\end{eqnarray}
since the point $x_1$ is located after the WKB invalid region.
Thus $x_1$ should be large enough so that the large $x$ expansion hold,
but it should not be too large to satisfy the above two relations.
The frequency squared for $\varepsilon x \gg  1$ becomes
$$
\Omega^2=\varepsilon^2\left(\bar r^2 + \frac{\bar q^2}{e^{2\varepsilon x}}\right)+ V_1(x), \quad \mbox{for } \varepsilon x\gg 1 .
$$
where
$$
\bar r = \frac{2^{2/3} kr}{3 H_i}, \quad
\bar q=\frac{2^{2/3} k}{3 H_i},\quad
V_1 =\frac{4\varepsilon^2}{3} \frac{\bar q^2}{e^{4\varepsilon x}}+\cdots.
$$

In this section, we consider the solution only to zeroth order.
The corresponding zeroth order solution for Eq.~(\ref{eom}) is given by the Bessel function:
\begin{eqnarray} \label{phi11}
\phi_0= \sqrt{\frac{\pi }{6H_i \sinh(\pi \bar r)}} \, J_{-i \bar r}(\bar q e^{-\varepsilon x}),
\quad  \mbox{for } \varepsilon x \gg 1.
\end{eqnarray}

Here we choose the solution so that it becomes an incoming wave form:
\begin{eqnarray}
\phi_{0}=\frac{1}{\sqrt{2 \cdot 2^{2/3} 
k_3}} \exp
\Big\{
i \bar{r}\varepsilon x
+i\psi \Big\}, \quad  \mbox{for } 
x\to \infty,
\end{eqnarray}
where the initial phase becomes
$e^{i \psi} =\left[
\frac{ \Gamma(1+i \bar r)}{\Gamma(1-i \bar r)}\right]^{1/2}
\left(\frac{\bar q}2\right)^{-i\bar r}$.
The accuracy of this solution is given by the ratio of the first order solution
to the zeroth order one.

Now we evaluate the accuracy at $x=x_1$
when the WKB approximation becomes valid again.
Green's function is given by
\begin{eqnarray}
G_{\rm ini}(x,y) = \frac{3 H_i}{i\varepsilon} [\phi_0(x) \phi_0^\ast(y)\theta(y-x)
    +\phi_0(y)\phi_0^\ast(x)\theta(x-y)].
\end{eqnarray}
Noting that $
\bar q e^{-\varepsilon x_1} \gg 1$ because of Eq.~\eqref{x1:kh},
the error can be estimated to be
$$
E_{\rm ini}=\left|\frac{\phi_1}{\phi_0}\right| =
    \left|\frac1{\phi_0} \int G_{\rm ini}(x,y) V_1(y) \phi_0(y) dy  \right|
  \sim 
\frac{k}{H_i} e^{-3 \varepsilon x_1}.
$$
Let us determine the cutoff time $x_1$
so that both of the solution~(\ref{phi11}) and the WKB solution are equally valid.
The WKB solution is valid if $\epsilon < 1$.
For $\varepsilon x \gg 1$, the accuracy of the WKB approximation is
\beq
E_{WKB}=\epsilon^2\approx \left(\frac{3}{2^{2/3}}\frac{H}{k}
    \frac{e^{-2\varepsilon x}}{(r^2 +e^{-2\varepsilon x})^{3/2}}\right)^2
\simeq
\left(\frac{H_i}{k}\right)^2e^{2\varepsilon x} \,,\nn
\eeq
where we use Eq.~\eqref{eex:r}.
The WKB approximation is valid during
 $$
  1\ll  e^{\varepsilon x} <
{\rm min} \big(\frac{k}{H_i},  r^{-1}\big) .
 $$
Both approximations are valid for this region.
Setting the two errors be of the same order, we choose $x=x_1$ to be
$$
e^{\varepsilon x_1} =c_1\left(\frac{k}{H_i}\right)^{3/5} ,
$$
where $c_1$ is an $O(1)$ constant.
It shows that the time $x_1$ is dependent on momentum but independent of directions.
Note that this value $e^{\varepsilon x_1}$ should be smaller than $r^{-1}$ because of \eqref{eex:r}.
Therefore, we have constraint for $r$:
$$
r\ll (H_i/k)^{3/5}.
$$
The accuracy at $x=x_1$ is given by $O(H_i/k)^{4/5}$.

During $x_\ast< x< x_1$, we use the WKB approximation where $x_\ast$ is given in Eq.~(\ref{t_ast}).
The WKB solution is
\beq \label{phi2}
\psi= \frac{B_+}{\sqrt{2\Omega}} \exp \Big\{i \int_{x_{1}}^{x} dx' \Omega(x')\Big\}
    +\frac{B_-}{\sqrt{2\Omega}} \exp \Big\{-i \int_{x_{1}}^{x} dx' \Omega(x')\Big\}.
\eeq
At $x=x_{1}$, to the zeroth order in $ \epsilon$, the matching condition becomes
\begin{eqnarray}
B_+ &=& \sqrt{\frac{\Omega_1}{2}} \left[\phi_0(t_1)-i\frac{\phi_0'(t_1)}{\Omega_1}\right],
\quad
 B_- =  \sqrt{\frac{\Omega_1}{2}}
\left[\phi_0(t_1)+i\frac{\phi_0'(t_1)}{\Omega_1}\right]
,
\end{eqnarray}
where
$$
\Omega_1^2 \simeq \varepsilon^2(\bar r^2 +\frac{\bar q^2}{e^{2\varepsilon x_1}})
 =\varepsilon^2 \frac{2^{4/3} k^2}{9H_i^2} \left(r^2 +\frac1{c_1^2}\left(\frac{H_i}{k}\right)^{6/5}\right)
 \simeq \frac{2^{4/3}}{9c_1^2} \left(\frac{H_i}{k}\right)^{11/5}.
$$
The argument of the Bessel function satisfies $\bar q e^{-\varepsilon x_1}\simeq \frac{2^{2/3}}{3} \left(\frac{k}{H_i}\right)^{2/5} \gg 1$.
Therefore, at $x=x_1$, we may employ the asymptotic form for the Bessel function:
\begin{eqnarray*}
\lim_{x\to \infty}J_{ia}(x) = \sqrt{\frac{2}{\pi x}} \cos\left(x-\frac{\pi}4-\frac{ia\pi}{2}\right) .
\end{eqnarray*}
and find
\begin{eqnarray} \label{phi1}
\phi_0= \sqrt{\frac{\pi }{6H_i \sinh(\pi \bar r)}} \,\sqrt{\frac{2}{\pi \bar q e^{-\varepsilon x_1} }}
   \cos (
    \bar q e^{-\varepsilon x_1} -\frac{\pi}{4}+\frac{i \bar r \pi}2   ) .
\end{eqnarray}
The coefficients $B_+$ and $B_-$  become
\begin{eqnarray} \label{B+-}
B_+ \simeq \sqrt{\frac{\varepsilon}{3
H_i}}\frac{e^{i (\frac{\pi}4-\bar q e^{-\varepsilon x_1})}}{\sqrt{1-e^{-2\pi \bar r}}}
 , \quad
B_- \simeq  \sqrt{\frac{\varepsilon}{3
H_i}}\frac{e^{-\pi \bar r}\, e^{-i (\frac{\pi}4-\bar q e^{-\varepsilon x_1})}}{\sqrt{1-e^{-2\pi \bar r}}}.
\end{eqnarray}
Note that $(H_i/k)^2 \ll \bar r \ll (k/H_i)^{2/5}$ and $B_\pm$ satisfy $|B_+|^2-|B_-|^2= \varepsilon/(3H_i)$.

For $x\ll x_{\ast}$, the WKB approximation is not valid.
However, for $x<x_\ast$ the asymptotic solution~(\ref{phi:fin}) works.
For $0<x<x_\ast$, the isotropic solution is valid and
\beq
\phi_0=A_+ u_0(x) + A_- v_0(x)\,.
\eeq

Matching the asymptotic solution with the WKB solution similarly to the previous section, we obtain
\begin{eqnarray}
\phi(x\to 0)
=-\big(A_++A_-\big)
&\simeq &
\Big(\frac{H_i^2}{2k^3}\Big)^{\frac{1}{2}}
\left(e^{-i \Psi}- e^{-\pi \bar r} e^{i \Psi}\right),
\end{eqnarray}
where $\Psi(k)=\bar k x_\ast^{1/3}-\int_{x_1}^{x_\ast} \Omega(x)dx+\bar q e^{-\varepsilon x_1} -\frac{\pi}4$.
This leads to the power spectrum of the form:
\begin{eqnarray} \label{P:planar}
P =
\Big(\frac{H_i}{2\pi}\Big)^2
\left(\coth \pi \bar r-\frac{\cos \big(2\Psi\big)}{\sinh \pi \bar r} \right).
\end{eqnarray}
The explicit value of the correction term becomes order of $10^{-3}$ for $\bar r\sim 3$.

At a glance,
the phase factor
$\Psi$
looks sensitive to
the choice of
the matching points $x_\ast$ and $x_1$ explicitly.
However,
it can be
confirmed that it is 
independent of them. 
Now, we compare two sets of the matching points
$(x_1,x_\ast)$ and $({\bar x}_1,{\bar x}_{\ast})$,
and take their difference to obtain 
\begin{eqnarray*}
\Psi-\bar{\Psi}
 &=&
{\bar k}(x_\ast^{1/3}-\bar{x}_{\ast}^{\frac{1}{3}})
+ \bar q 
   \big(e^{-\varepsilon x_1 }-e^{-\varepsilon \bar{x}_1 }\big)
-\int^{x_\ast}_{{\bar x}_\ast} \Omega(x) dx
+\int^{x_1}_{{\bar x}_1} \Omega(x) dx
\nonumber\\
&\simeq &
{\bar k}(x_\ast^{1/3}-\bar{x}_{\ast}^{\frac{1}{3}})
+ \bar q 
   \big(e^{-\varepsilon x_1 }-e^{-\varepsilon \bar{x}_1 }\big)
-\int^{x_\ast}_{{\bar x}_\ast} 
     \frac{\bar k}{3x^{2/3}}
    dx
+\int^{x_1}_{{\bar x}_1} 
  {\bar q}\varepsilon e^{-\varepsilon x}
 dx
=0,
\end{eqnarray*}
where we have used the values of the frequency squared around $x=x_\ast$ and $x=x_1$.

After averaging the power spectrum (\ref{P:planar}) over the plane,
i.e., integrating over the momenta $k_1$ and $k_2$,
the second term of the spectrum
vanishes
and
the main contribution
comes from the first term, which is approximated as
\beq
\bar{P}
\simeq
\left(\frac{H_i}{2\pi}\right)^2
\coth (\pi \bar r)
= \left(\frac{H_i}{2\pi}\right)^2
\coth \left(
\frac{2^{2/3}\pi k}{3H_i}\cos\theta
\right),
\label{ave}
\eeq
where we have used the definition of $\bar r$ and 
$r=\cos\theta$.
Thus in the planar limit,
the deviation of the power spectrum from the ansatz in Ref. \cite{ACW}
is quite clear.
It makes it impossible to compare observations
through the analysis based on the ansatz in \cite{ACW,pullen,gron,ma}.

\subsection{
Physical bounds for $r$
}

Our analysis in this paper
indicates that
the power of the scalar field fluctuation has a negligible
 difference from the standard inflation in the non-planar directions,
but it has a sharp peak around the plane $\theta=\pi/2$.
In other words, the power is strongly localized around the plane $\theta=\pi/2$.
While such a feature is much different from the often used phenomenological ansatz
for the anisotropic power spectrum Refs. \cite{ACW, pullen},
it also implies a band-like structure in the CMB temperature map,
around which the averaged temperature is much different from that in the other direction.
Recently, it was reported that giant ring structures which have an anomalous mean temperature profile were
observed in the CMB sky \cite{gr1,gr2}.
Thus it would be very interesting
to compare the predictions in our model with
such a new observational feature.

In the rest of this section, we
discuss the possible bound on the parameter.
In particular,
while the power spectrum in the planar mode Eq. (\ref{P:planar})
diverges in the $r=0$ limit,
we also find the purely planar mode $r=0$ cannot be quantized.
It implies that we need to introduce a cut-off
on the $r$
to be consistent with theoretical and/or observational
requirements.
From the above analysis,
we have found that
for a large $\bar r> 2$, the power spectrum~\eqref{P:planar} is not much different
from that of the de Sitter
(the difference is too small; the difference is $O(10^{-5})$ for $\bar r=2$).
Thus we focus on the case $\bar r <2$.

A bound is obtained from the validity of our method.
The present approximation holds for $(k/H_i)^{2/5}\gg\bar r \gg H_i^2/k^2$.
The amplification will be maximized for a small $\bar r$.
By defintion of $\bar r$, for a given $k$,
it is equivalent to put the bound on $r$
as $(k/H_i)^{-3/5}\gg r \gg H_i^3/k^3$.
This bound is from the validity of our approximation.
But it can only give a $k$-dependent bound on $r$.
However,
since $k$ and $r$ are independent variable in the Fourier space,
a $k$-independent bound may be obtained
from the consistency with requirements of theory and observation.
Now, we would like to check whether our scalar field solution
naturally satisfies the required bound or not.

From the theory side,
we may check the backreaction of the scalar field
may modify the background evolution
in the infinite past
where the scalar field is quantized
in the anisotropic vacuum.
Since the UV divergence may be cured by the standard renormalization,
here we focus on the IR behavior
(For UV, here we simply put the UV cut-off $k_{uv}$,
which may be chosen close to the Planck scale $m_p$).
Now we check the sensitivity of the scalar field backreaction
to the IR cut-off $k_{ir}$ for $k$-integration and
the fiducial cut-off $0<r_0\ll 1$ in the $r$-integration.
If the energy density is
very sensitive, it would lead a bound on $k_{ir}$ and $r_0$.
In the Appendix~\ref{app:d}
we explicitly compute the energy density of the scalar field
$\rho$ after introducing the ultra-violet and infra-red cutoff
and the leading order result in the infinite past is given by
$$
\rho(X\to 0)
\simeq
    \frac{\pi  X^{-1}}{2^{1/3}(2\pi)^3}
    \int_{k_{ir}}^{k_{uv}} dk k^3
   \int_{r_0}^1 dr  r,
$$
where $X= e^{-2x}$.
As seen here, there is no infra-red divergence, leaving no bound on the IR part
and one may naturally take $r_0\to 0$ and $k_{ir}\to 0 $ limit.
\footnote{For a different point of view, see~\cite{Paban2}.}

Finally, we turn to the observational side.
One of the most important requirements from observations
is that the anisotropic contribution on the power spectrum
should not exceed the isotropic one.
In the Appendix \ref{app:e},
we have computed the corrections
in the CMB temperature anisotropy.
Note that although in this paper
we have not discussed the metric perturbations,
the previous works show
that our scalar field perturbations
share the same features with
the metric perturbations \cite{km,km2}
and here we proceed with the power spectrum obtained for the scalar field.
After some computations,
for a given $\ell$,
the ratio of the corrections from
the anisotropy $\Delta C_{\ell\ell m}$
to the standard isotropic
angular multipole moment is given by
\bea
\frac{\Delta C_{\ell\ell m}}{C_\ell}
&\sim&
{\tilde C} (\ell,m)
\left[(
k_{ir} \eta_\ast) \ln r_0^{-1}\right],
\eea
where
$\eta_\ast$ is the comoving distance
from the last scattering surface to the present time
and
${\tilde C} (\ell,m)$ is given by Eq. (\ref{clm}).
Note that the above result is valid
when $\ell+m$ is even.
Avoiding the large anisotropic contribution,
$\Delta C_{\ell\ell m}/C_\ell<1$,
gives
a lower bound on $r_0$
for each $\ell$ and $m$:
\bea
r_0> {\rm exp}\Big[
-\frac{1}{{\tilde C}(\ell,m) (
k_{ir} r_\ast)}
\Big].
\eea
Among the bounds obtained for various $\ell$,
we should choose the strongest bound.
For $m=0$, for example,
${\tilde C}(2,0)\simeq 0.267$,
${\tilde C}(10,0)\simeq 0.0576$,
${\tilde C}(20,0)\simeq 0.0284$,
and hence for a larger even $\ell$,
${\tilde C}(\ell,0)$ becomes smaller.
Thus, we should impose a bound on $r_0$
from one of the lowest $\ell$:
focusing on $\ell=2$,
\bea
\label{dmi}
r_0>  {\rm exp}\Big[
-\frac{3.70}{k_{ir} r_\ast}
\Big].
\eea
If we put $k_{ir} \eta_\ast=1$,
the right-hand side of Eq. (\ref{dmi})
becomes $0.0246$,
which is reasonably small.
As shown in the Appendix. \ref{app:e},
$H_i \eta_\ast\gtrsim 2$.
Choosing the IR cut-off $k_{ir}\lesssim H_i$,
the above lower bound can be naturally obtained.
In other words, the lower bound can be made less serious because we may choose the parameter $k_{ir}$ freely.
If the giant ring or any other specific effect would be observed in the future,
it would finally determine the value of $r_0$.

\section{Conclusion} 

In this article, we have reinvestigated the quantization of a massless and minimally coupled scalar field as
a way to probe the signature of pre-inflationary background anisotropy in the spectrum of cosmological perturbations.
We have considered the gravitational theory composed of the Einstein-Hilbert term and a positive cosmological constant.
For simplicity, we have assumed that the late-time inflationary stage is exactly described by the de Sitter solution and
ignored the dynamics of
the inflaton field, which can be justified by the slow-roll approximations as shown in Ref.~\cite{km2}.
The initial geometry we have considered is the axially symmetric Bianchi I metric which evolves
to the de Sitter spacetime due to the positive cosmological constant.
The motivation to consider the initially anisotropic Universe is two-fold:
The first is that even if the current Universe is almost isotropic, it does not mean that the Universe is isotropic from the beginning.
In fact, Wald's no-hair conjecture ensures that in the presence of a positive cosmological constant an initially anisotropic Universe exponentially
approaches a de Sitter spacetime at the later time under the strong or dominant energy condition.
Therefore, it would be more generic that the initial Universe is anisotropic and
one expects that remaining initial anisotropy is imprinted in cosmic observables.
The second motivation comes from the recent observations by 
WMAP satellite which almost confirmed the predictions from 
inflation, during which the Gaussian and statistically isotropic primordial fluctuations are produced.
But, several groups have reported the possibility to detect the remnants of the preinflationary anisotropy through the fluctuation spectrum~(See \cite{Copi:2010na} and references therein).

We first dealt the non-planar modes.
We have divided the time into two periods, before and after the mode-dependent time $\tau_\ast \simeq \frac1{H_{i}}\log\frac{k}{{H_{i}}}$.
We used the WKB approximation for the first part $(\tau < \tau_\ast)$ and used the asymptotic approximation based on the de Sitter solution for the next part $(\tau > \tau_\ast)$.
At $\tau=\tau_\ast$, the two approximations have the same accuracy with an error of order ${H_{i}}/k$.
We illustrated the calculation up to the order of $({H_{i}}/k)^3$ in Sec.~\ref{sec:3C} and up to the order of $({H_{i}}/k)^6$ in Appendices A to C.
We have shown that,
at each order, the asymptotic value of the scalar field acquires corrections from that of the de Sitter form and has the form,
$$
\phi_{\rm fin\, \bf k} = (A_{\bf k}+ B_{\bf k} i)e^{-i \sqrt{k/H_i}},
$$
where $A_{\bf k}$ and $B_{\bf k}$ are real constants which are dependent on the momentum.
Since the phase part 
is written as a multiplicative factor,
$|\phi_{\rm fin}|^2= A_{\bf k}^2+B_{\bf k}^2$ gains no momentum dependent oscillatory behavior.
It was also shown that the power spectrum of the scalar field acquires non-vanishing corrections only when we execute the approximation up to $6^{\rm th}$ order.
Hence, the direction dependence appears only at the order $O(({H_{i}}/k)^6)$.
If we want to observe the direction dependence from the power spectrum (other than the planar mode), the wavelength of the fluctuation due to the modes with $H_{i}<k_{\rm crit}< k<10H_{i}$ should be located in the observable range of the universe $(1-10^4)\mbox{Mpc}$.
For modes with $k> 10 H_{i}$, we can not detect the effect with the present equipment because it is too small (the fluctuation will be smaller than $<10^{-6}$).
For modes with $k < H_{i}$, the approximation we adopted in this paper is unapplicable and we cannot apply the result~(\ref{P:phi}).
In addition, it also restrict $k_{\rm crit}< 10 H_{i}$, which gives $N< 64+\log_e(10)\simeq 66$ in the case of GUT scale inflation.
In summary, if the power spectrum is observed in the previous form, it restricts the number of $e$-folds to the narrow range $64< N<66$.
If the number of $e$-folds is larger than $66$, we may claim that the effect of anisotropy will be unobservable within the present accuracies of
observations.
If the number of $e$-folds is smaller than $64$, some of long wavelength fluctuations,
which 
are of the order of the scale of our universe, cannot be dealt with the present adiabatic approximation.
The two-point function in this regime was numerically analyzed in~\cite{gcp}.

For the planar mode, we have obtained essentially
the same result as that in our previous analysis \cite{km,km2}.
For such a mode,
the temporal breaking of the WKB approximation relatively enhances the
effects of the primordial anisotropy in the power spectrum.
For $\frac{k_3}{H_i}\sim 3$, the deviation from the ordinary case
becomes of order $10^{-3}$.
We also discussed the observability of the planar mode contributions.
It was shown that there exists a sharp peak around the symmetric plane.
The giant ring structure in the CMB sky map can be related with this result.
Note that our result is applicable even for a small $k_3\sim H_i/10 $.
In this case, the power spectrum increases by a factor $6$ to the standard de Sitter case, which was not observed.
There are three possibilities to cure this defect.
The first is that there may exist a lower bound on the momentum orthogonal to the plane, which was discussed in Sec IV B.
The second possibility is that the large energy density of the scalar field reacts to change the background geometry.
The third possibility is that the gravity theory can be modified at the ultra high energy scale so that the anisotropy may not grow indefinitely at initial times as in the case of the Eddington-Inspired gravity~\cite{Eddington}.

The properties of the non-planar mode function
also suggest that it would be very difficult
to obtain distinguishable amounts of the primordial non-Gaussianities
in the present model.
The bispectrum is very sensitive
to physical process before, during and after inflation.
Assuming three comoving momenta ${\bf k}_i$ ($i=1,2,3$)
which forms a triangle in the Fourier space, $\sum_i {\bf k}_i=0$,
interactions
just before and after the horizon crossing (i.e., classical interactions)
give rise to the equilateral-type ($k_1\simeq k_2 \simeq k_3$
where $k_i:=|{\bf k}_i|$)
and local-type ($k_1\ll {\rm min}(k_2,k_3)$ and permutations) bispectra,
respectively.
On the other hand,
the existence of an excited state 
in the past
contributes to the folded-type bispectrum ($k_1+k_2\simeq k_3$,
and permutations) \cite{nG:study}.
Since after the isotropization
the evolution is the same as the conventional single-field slow-roll inflation,
only the possible origin of the distinguishable non-Gaussianities may be
the folded-type one
due to the modification to the anisotropic vacuum.
The absence of the negative frequency mode,
however, implies that a negligible difference from the standard case
for the non-planar high-momentum modes.

For the planar mode,
since the contribution of the negative frequency mode
is less suppressed than for the non-planar mode,
the amplitude of the folded-type bispectrum
may be nontrivially different from that of the standard case
and have features of modulations.
In the present model, however,
the overall amount of the primordial non-Gaussianity is
already suppressed by the slow-roll parameters,
and hence it would be still very difficult
to detect their difference with the precision of the near future observations
such as the Planck satellite.
Non-Gaussianities may be enhanced
in the inflationary models with noncanonical kinetic terms for the inflaton fields.
It would be interesting to observe the process of the isotropization
in such a generalized inflation model.

\section*{Acknowledgement}
HCK was supported in part by the Korea Science and Engineering Foundation
(KOSEF) grant funded by the Korea government (MEST) (No.2010-0011308) and the APCTP Topical Research Program (2012-T-01).
MM is supported by Yukawa fellowship and by Grant-in-Aid for Young Scientists (B) of JSPS Research, under Contract No. 24740162.
MM is also grateful to the School of Liberal Arts and Sciences, Korea National University of Transportation,
for the hospitality.

\appendix

\section{Higher order WKB approximation at early times}\label{app:A}

We need to solve Eq.~(\ref{WKB:iteration}) iteratively.
To zeroth order, the frequency squared is given by
$$
\tilde\Omega_0^2 = \Omega^2.
$$
Including the first order correction, it is given by
$$
\tilde\Omega_1^2
     = \Omega^2\left[1 -\frac{\tilde \Omega_{0,xx}}{2\Omega^2\tilde \Omega_0}
        +\frac34\left(\frac{\tilde \Omega_{0,x}}{\Omega\tilde \Omega_0}\right)^2\right]
     =\Omega^2\left[1 -\frac{\Omega_{,xx}}{2\Omega^3}
        +\frac34\left(\frac{\Omega_{,x}}{\Omega^2}\right)^2\right],
$$
which is correct to the order of $\varepsilon^{n+1}$.
Note that the first order correction terms are of 
order $O(\varepsilon)$.
In this way, the higher order corrected frequency squared is given by the iteration,
$$
\tilde\Omega_{n+1}^2
     = \Omega^2\left[1 -\frac{\tilde \Omega_{n,xx}}{2\Omega^2\tilde \Omega_n}
        +\frac34\left(\frac{\tilde \Omega_{n,x}}{\Omega\tilde \Omega_n}\right)^2\right].
$$

To have accuracy up to $O(\varepsilon^{5})$,
we need to get $\tilde \Omega^2_5$ iteratively.
For $\varepsilon x\ll 1$,
through the successive iterations,
we obtain
\begin{eqnarray}
\left[\frac{\tilde\Omega}{\Omega}\right]_{x=1} &=&
1-\frac{1}{{\bar k}^2}+\frac{1}{{\bar k}^4}+\frac{3 r^2-1}{6 {\bar k}^5} -\frac{1}{{\bar k}^6}+\frac{\frac{1}{3}-r^2}{{\bar k}^7}+\frac{549 r^4-498 r^2+133}{72 {\bar k}^8}+\frac{3
   r^2-1}{2 {\bar k}^9}+\frac{-45 r^4+41 r^2-8}{3 {\bar k}^{10}}\nn \\
&+&\frac{-14175 r^6+19629 r^4-7191 r^2+485}{324 {\bar k}^{11}}+\frac{-549 r^4+486 r^2+35}{96 {\bar k}^{12}}+\cdots.
\end{eqnarray}
In addition, we 
obtain
\begin{eqnarray}
\left[\frac{\tilde\Omega'(x)}{\tilde \Omega(x)^2}\right]_{x=1} &=&
    -\frac{2}{{\bar k}} \left[1+\frac{5 q^2}{2 {\bar k}^3}+\frac{9 q^2}{4 {\bar k}^5}+\frac{6 q^4+\frac{4 q^2}{3}-\frac{4}{9}}{{\bar k}^6}-\frac{3 \left(45 q^4+12 q^2-4\right)}{8 {\bar k}^8}+\frac{11 \left(405 q^6+162 q^4-54 q^2-8\right)}{324 {\bar k}^9}\right. \nn\\
&-&\left.\frac{5 \left(2079 q^6+756
   q^4-252 q^2-32\right)}{48 {\bar k}^{11}}
   +O\left(\frac{1}{{\bar k}^{12}}\right)
   \right].
\end{eqnarray}

\section{Perturbative calculations in the later times}\label{app:B}

We have used the iterative perturbation method. 
To solve Eq.~(\ref{eom:series}), we first set the solution in series form as
$$
u = \sum_{n=0}^\infty \varepsilon^n u_n(x), \quad
v = \sum_{n=0}^\infty \varepsilon^n v_n(x),
$$
where $u$ and $v$ are two independent solutions of the equation of motion~(\ref{eom:series}) and are complex conjugates to each other.

The zeroth order solution is given in Eq.~(\ref{uv}).
Once we obtain the solution up to $n^{th}$ order,
$(n+1)^{th}$ order correction can be found,
by solving the following differential equation,
\begin{eqnarray}
&&\left(\frac{d^2}{dx^2} +\frac{\bar k^2}{9x^{4/3}}\right)u_{n+1} =
    -\sum_{m=0}^{n} u_{m}(y) V_{n-m}(y),
\end{eqnarray}
 with the source term becomes
\begin{eqnarray}
&&V_1=-\frac{2\bar k^2 q^2}{9 x^{1/3} }, \qquad
    V_2=\frac{2\bar k^2 r^2 x^{2/3}}{27} ,\qquad
    V_3= \frac{8 \bar k^2 x^{5/3}}{729},\qquad V_4=  -\frac{\bar k^2 (11 + 30 r^2) x^{8/3}}{10935}, \nn \\
&&V_5 =\frac{2 \bar k^2 \left(r^2-7\right) x^{11/3} }{10935}, \qquad
    V_6= \frac{2 \bar k^2 \left(189 r^2+170\right) x^{14/3}}{2066715},\qquad
    \cdots .  \label{Vi}
\end{eqnarray}

Therefore, the $(n+1)^{th}$ order solution can be obtained,
by using the Green's function~(\ref{greenfn}), as
$$
u_{n+1}(x) = -\int dy \, G(x,y) \left[\sum_{m=0}^{n} u_{m}(y) V_{n-m}(y)\right].
$$
By using this method, we get
\begin{eqnarray}
u_1(x) &=&e^{i \bar{k}  \sqrt[3]{x}} \left(\frac{1}{4} q^2 x^{5/3} \bar{k} ^2+\frac{3}{4} i q^2 x^{4/3} \bar{k} -\frac{3}{4} \left(q^2 x\right)-\frac{9 \left(q^2 \sqrt[3]{x}\right)}{8 \bar{k} ^2}-\frac{9 i q^2}{8
   \bar{k} ^3}\right)+e^{-i \bar{k}  \sqrt[3]{x}} \left(-\frac{9 \left(q^2 \sqrt[3]{x}\right)}{8 \bar{k} ^2}+\frac{9 i q^2}{8 \bar{k} ^3}\right), \nn\\
u_2(x)&=&e^{-i \bar{k}  \sqrt[3]{x}} \left[-\frac{9 i q^4 x^{5/3}}{32 \bar{k} }
      -\frac{27 \left(q^4 x^{4/3}\right)}{32 \bar{k} ^2}+\frac{27 i q^4 x}{32 \bar{k} ^3}+\frac{9 i \left(99 q^4+24 q^2-8\right) \sqrt[3]{x}}{64 \bar{k} ^5}+\frac{9 \left(99 q^4+24 q^2-8\right)}{64 \bar{k} ^6}\right]\nn \\
&+&e^{i \bar{k}  \sqrt[3]{x}} \Big[-\frac{1}{32} i q^4 x^3 \bar{k} ^3+\frac{\left(459 q^4+96 q^2-32\right) x^{8/3} \bar{k} ^2}{2016}+\frac{1}{112} i \left(99 q^4+24 q^2-8\right) x^{7/3} \bar{k} \nn \\
&-&\frac{1}{40} \left(99 q^4+24 q^2-8\right) x^2-\frac{9 i \left(47 q^4+12 q^2-4\right) x^{5/3}}{80 \bar{k} }+\frac{3 \left(45 q^4+12 q^2-4\right) x^{4/3}}{16 \bar{k} ^2}+\frac{3 i \left(45 q^4+12 q^2-4\right) x}{16 \bar{k} ^3}\nn \\
&+&\frac{9 i \left(99 q^4+24 q^2-8\right) \sqrt[3]{x}}{64 \bar{k} ^5}-\frac{9 \left(99 q^4+24 q^2-8\right)}{64 \bar{k} ^6}+\cdots \Big],
\end{eqnarray}
\begin{eqnarray}
u_3(x) &=&  e^{-i \bar{k}  \sqrt[3]{x}} \left[\frac{9 q^6 x^3}{256}
     -\frac{i q^2 \left(459 q^4+96 q^2-32\right) x^{8/3}}{1792 \bar{k} }-\frac{9 \left(q^2 \left(99 q^4+24 q^2-8\right) x^{7/3}\right)}{896 \bar{k} ^2}+\frac{9 i q^2 \left(99 q^4+24 q^2-8\right) x^2}{320 \bar{k} ^3}\right] \nn \\
&+& e^{i \bar{k}  \sqrt[3]{x}} \Big[-\frac{1}{384} \left(q^6 x^{13/3}\right) \bar{k} ^4
      -\frac{i q^2 \left(291 q^4+96 q^2-32\right) x^4 \bar{k} ^3}{8064}+\frac{\left(102789 q^6+40176 q^4-13392 q^2-1792\right) x^{11/3} \bar{k} ^2}{362880}\nn \\
&+&\frac{i \left(50193 q^6+19224 q^4-6408 q^2-896\right)
     x^{10/3} \bar{k} }{30240}+\frac{\left(-273807 q^6-101952 q^4+33984 q^2+4480\right) x^3}{34560}\nn \\
&-&\frac{i \left(284607 q^6+104352 q^4-34784 q^2-4480\right) x^{8/3}}
     {8960 \bar{k} }+\frac{\left(288387 q^6+105192 q^4-35064 q^2-4480\right) x^{7/3}}{2688 \bar{k} ^2}\nn \\
&+&\frac{i \left(288387 q^6+105192 q^4-35064 q^2-4480\right) x^2}
     {960 \bar{k} ^3}\Big]+\cdots,
\end{eqnarray}
\begin{eqnarray}
u_4(x) &=&e^{-i \bar{k}  \sqrt[3]{x}}
     \left[\frac{3 i q^8 x^{13/3} \bar{k} }{1024}+\frac{q^4 \left(291 q^4+96 q^2-32\right) x^4}{7168}\right]
  +e^{i \bar{k}  \sqrt[3]{x}} \left[\frac{i q^8 x^{17/3} \bar{k} ^5}{6144}
     - \frac{\left(q^4 \left(477 q^4+192 q^2-64\right) x^{16/3}\right) \bar{k} ^4}{129024} \right. \nn \\
&-&\frac{i \left(484623 q^8+247752 q^6-71064 q^4-20224 q^2+1280\right)
    x^5 \bar{k} ^3}{10160640} \nn \\
&+&\frac{\left(30240459 q^8+15875352 q^6-4410504 q^4
      -1553408 q^2+141824\right) x^{14/3} \bar{k} ^2}{66044160}\nn \\
&+&\frac{i \left(5658255 q^8+2920077 q^6-812511 q^4-278256 q^2+25712\right) x^{13/3} \bar{k} }{1572480} \nn \\
&-&\left. \frac{\left(64266939 q^8+32444658 q^6-9097542 q^4-2976096 q^2+269216\right) x^4}{2661120}\right]+\cdots,
\end{eqnarray}
\begin{eqnarray}
u_5(x)&=& e^{i \bar{k}  \sqrt[3]{x}} \Big[\frac{q^{10} x^7 \bar{k} ^6}{122880}+\frac{i q^6 \left(2133 q^4+960 q^2-320\right) x^{20/3} \bar{k} ^5}{7741440}\nn \\
&-&\frac{\left(q^2 \left(853821 q^8+519264 q^6-127008 q^4-55808 q^2+5120\right) x^{19/3}\right) \bar{k} ^4}{162570240}\nn \\
&-&\frac{i \left(175688271 q^{10}+114725160 q^8-25188408 q^6-16032384 q^4+1286400 q^2+186368\right) x^6 \bar{k} ^3}{2377589760}\Big]+\cdots,
\end{eqnarray}
\begin{eqnarray}
u_6(x)&=& e^{i \bar{k}  \sqrt[3]{x}} \left[-\frac{i q^{12} x^{25/3} \bar{k} ^7}{2949120}+\frac{q^8 \left(993 q^4+480 q^2-160\right) x^8 \bar{k} ^6}{61931520}\right]+\cdots ,
\end{eqnarray}
The independent solution $v(x)$ is nothing but the complex conjugate of $u(x)$.
The Wronskian can also be checked 
to give
\begin{eqnarray} \label{wronskian:uv}
u v'-u'v = -\frac{2i{\bar k}^3}{3} +O({\bar k}^{-10}).
\end{eqnarray}

\section{The asymptotic form of the scalar field}\label{app:C}

In the limit of the later times $x\to 0$, the wavefunction behaves as
\beq
 \phi_{\rm fin}\Big|_{x\to 0}
&=&-(A_+ + A_-)= \frac{3i}{2\bar k^3}\left[\phi_{\rm WKB}(1)(u'(1)-v'(1))-
     \phi_{\rm WKB}'(1)(u(1)-v(1))\right]\nn \\
&=&i \sqrt{\frac{3\varepsilon}{2{H_{i}} \bar k^3}} \frac{\Phi_\ast}{\sqrt{2}}
    \left\{\sqrt{\frac{\bar k}{\tilde \Omega_*}}\frac{u'(1) - v'(1)}{\bar k^2}-
        i\left(1+\frac{i\tilde\Omega'(x_\ast)}{ 2\tilde\Omega_*^2} \right) \sqrt{\frac{\tilde\Omega}{\bar k}}\, \frac{ u(1)-v(1)}{\bar k} \right\}
. \nn
\eeq
By using the perturbative results in the previous Appendices~\ref{app:A} and \ref{app:B}, we get
\begin{eqnarray*}
&&\sqrt{\frac{2{H_{i}} \bar k^3}{\varepsilon}} \frac{e^{i \bar k}}{\Phi_\ast} \phi_{\rm fin}
=i+\frac{1}{{\bar k}}+\frac{-\frac{q^2}{4}-\frac{i}{2}}
   {{\bar k}^2}+\frac{-\frac{1}{2}+\frac{i q^2}{4}}{{\bar k}^3}-\frac{i \left(q^4+12 i q^2-12\right)}{32 {\bar k}^4}\nn \\
&&\quad+\frac{-207 q^4-(96+756 i)
   q^2+788}{2016 {\bar k}^5}+\frac{21 q^6+1206 i q^4-(3780-384 i) q^2-2648 i}{8064 {\bar k}^6}  \nn \\
&&\quad-\frac{i \left(825 q^6+(480+74142 i) q^4-(19060-18432 i) q^2-18744 i\right)}{40320
   {\bar k}^7}  \nn \\
&&\quad+\frac{i \left(945 q^8+446904 i q^6-(11016648-262656 i) q^4
    -(2654208+3262752 i) q^2+(2472336-28672 i)\right)}{5806080 {\bar k}^8} \nn \\
&&\quad +\frac{13905 q^8+(8640+3002616 i)
   q^6-(37893384-926208 i) q^4-(9103104+3483936 i) q^2+(4621968-28672 i)}{5806080 {\bar k}^9}\nn \\
&&\quad+\frac{1}{162570240 {\bar k}^{10}}\Big[-1323 q^{10}-2928798 i q^8
    +(831208392-2149632 i) q^6+(338204160+1072579536 i)
   q^4\nn \\
 &&\quad -(212753520-255210496 i) q^2-(17461248+124990304 i)\Big]\nn \\
&&\quad +\frac{i}{2113413120 {\bar k}^{11}}\Big[ 410319 q^{10}+(262080+260683110 i) q^8
    -(14181596520-142566912 i)
   q^6 \nn \\
&&\quad-(5225036544-163571184 i) q^4+(3041923248+133211136 i) q^2+(226996224+471614432 i)\Big] \nn \\
&-&\frac{i}{16907304960 {\bar k}^{12}} \Big[5733 q^{12}+41102100 i
   q^{10}-(23757489540-34085376 i) q^8-(11283840000+1401723182496 i) q^6\nn \\
&&-(232347906000+516373012480 i) q^4-(57447653376-183565086016 i) q^2+(15556834880+22417833984
   i)\Big]\nn \\
&&+O({\bar k}^{-13}).
\end{eqnarray*}
Notice that this is of the form $\phi_{\rm fin \, \bf k} = (A_{\bf k} + B_{\bf k} i)e^{-i {\bar k}}$ where $A_{\bf k}$ and $B_{\bf k}$ are real numbers.
Therefore, $|\phi_{\rm fin\, \bf k}|^2=A_{\bf k}^2+B_{\bf k}^2$ has no oscillatory behavior with ${\bar k}$.

\section{Quantum backreaction in the infinite past} \label{app:d}

In this Appendix, we discuss the quantum backreaction
of the scalar field in the infinite past.
We rewrite the metric Eq. (\ref{ds2:KdS})
into that written in the new coordinate $x$,
defined by
$\sinh(3H_i\tau)=\frac{1}{\sinh x}$.
In the $x$-frame,
the metric can be rewritten as
\bea
ds^2=-e^{6\alpha} \Big(\frac{1} {3H_i}\Big)^2 dx^2
+e^{2\alpha}
\Big[
 e^{-2\beta}(dx_1^2+dx_2^2)
+e^{4\beta} dx_3^2
\Big],
\eea
where
\bea
e^{\alpha(x)}=\frac{1}{\sinh^{1/3} x},
\quad
e^{\beta(x)}=e^{-\frac{1}{3}x}.
\eea

We are interested in the early time behavior of the backreaction and
adopting the leading order WKB approximation,
\bea
\label{wkb}
&&\phi_k \simeq \sqrt{\frac{1
}{3H_i }}
\frac{1}{\sqrt{2\Omega }}
{\rm exp} \Big\{ i \int dx' \Omega \Big\},\quad
\phi_{k,x} \simeq 
i{\Omega}
\phi_{k}.
\eea
Decomposing the scalar field into the Fourier modes
as Eq. (\ref{ds2:KdS}),
the energy density of the scalar field is given by
\bea
\rho=\frac{1}{2(2\pi)^3}
\int d^3k
\Big[
e^{-6\alpha} 
(3H_i)^2     \big| \phi_{k,x} \big|^2
+e^{-2\alpha}
   \Big(
    e^{2\beta}    k_{\perp}^2
+ e^{-4\beta}    k_3 ^2
   \Big)
 \big|\phi_{k}  \big|^2
\Big].
\eea
Substituting $\alpha(x)$, $\beta(x)$
and mode functions,
the energy density is rewritten
in terms of the integration of the Fourier mode
\bea
\rho(x)
=\frac{1}{2(2\pi)^3}
\int d^3k
\Big[
(3H_i)^2
\sinh^2 x
      \big|\phi_{k,x}  \big|^2
+ \sinh^{2/3}  x
  \Big(
    e^{-2
 x/3}    k_{\perp}^2
+ e^{4
 x/3}    k_3 ^2
   \Big)
 \big|\phi_{k}  \big|^2
\Big].
\label{four}
\eea

Substituting Eq. (\ref{wkb}) into Eq. (\ref{four}) we get
\bea
\rho
&=&
\frac{e^{2x}}{4(2\pi)^3}
\left(\frac{1}{3H_i}\right)
\int d^3k
\frac{1}{\Omega}
\Big[
(3H_i)^2
\Big(\frac{1}{4}\Big)
\big(1-e^{-2x}\big)^2
{\Omega^2}
+ k^2
 \Big(\frac{1}{4}\Big)^{1/3}
\big(1-e^{-2x}\big)^{2/3}
 \Big(
e^{-2x}  (1-r^2)
+  r^2
   \Big)
\Big].
\eea
For the further convenience,
introducing a new coordinate $X:= e^{-2x}$,
and using $d^3k = 4\pi k^2 dk dr$
(here, we have taken the symmetry between the north and south
hemispheres into account) 
we write the energy density as
\bea
\rho(X)
=\frac{\pi X^{-1}}{(2\pi)^3}
\Big(\frac{1}{3H_i }\Big)
\int_{k_{ir}}^{k_{uv}} dk k^2
\int_{r_0}^1 dr
U_k(X,r),
\label{nar}
\eea
where we explicitly introduced the ultra-violet and infra-red cutoffs explicitly.
Here,
\bea
U_k (X,r)&:=&
\frac{1}{\Omega}
\Big[
\frac{(3H_i)^2}{4}
\big(1-X\big)^2
{ \Omega^2}
+k^2
 \Big(\frac{1}{4}\Big)^{1/3}
\big(1-X\big)^{2/3}
  \big(
(1-r^2) X
+  r^2\big)
\Big].
\eea
Now the UV divergence is cured by the UV cut-off $k_{uv}$
which may be chosen close to the Planck scale $m_p$.
In the rest of this Appendix,
we focus on the IR divergence.

Following the definition
\bea
\Omega^2 (X)
=\left(\frac{k}{3H_i}\right)^2
\frac{2^{\frac{4}{3}} \left[ (1-r^2) X+r^2\right]}
     {(1-X)^{4/3}},
\eea
in the infinite past,
the energy density is given by
\bea
\rho(X\to 0)
\simeq
\frac{\pi  X^{-1}}{2^{1/3}(2\pi)^3}
\int_{k_{ir}}^{k_{uv}} dk k^3
\int_{r_0}^1 dr  r.
\eea
Hence there is no IR divergence
both in the $k$- and $r$- integrations.

\section{CMB anisotropy}~\label{app:e}

In this Appendix,
we evaluate the CMB temperature anisotropy $\frac{\delta T}{T}$ in our model.
The power spectrum of the primordial fluctuations
is in general given by
\bea
\langle
\delta ({\bf k})
\delta^{\ast} ({\bf k'})
\rangle
=(2\pi)^3 P({\bf k})\delta^{(3)}({\bf k}-{\bf k'}).
\eea
Here, we assume that
the power spectrum of the primordial perturbations
including the effects of the primordial anisotropy
is given  by
\bea
P(\bk)= P_0(k)\Big(1+ \g(k,\Omega_k)\Big),
\label{ansatz}
\eea
where $\Omega_k$ denotes the solid angle in the Fourier space
and $P_0(k)$ is the power spectrum in the isotropic case.
Note that we have not worked on the metric perturbations,
but following the results obtained in \cite{km,km2}
we can expect that
the form of $\g$ is not modified much from the case of the scalar field
even in cases of the metric perturbations.

The CMB temperature anisotropy can be expanded
in terms of the angular multipole moments
given by
\bea
a_{\ell m}= \int d{\Omega} Y_{\ell m}(\Omega) \frac{\delta T}{T}(\Omega),
\eea
where $Y_{\ell m} (\Omega)$
is the spherical harmonic function satisfying the normalization condition
\bea
\int d\Omega Y_{\ell m}(\Omega)Y_{\ell' m'} (\Omega)
= \delta_{\ell \ell'}\delta_{m m'}.
\label{nc}
\eea
The expectation value of the angular multipole moment is given by
\bea
\langle
a_{\ell m}
a^{\ast}_{\ell' m'}
\rangle
=\int d\Omega d\Omega'
\Big\langle
\frac{\delta T}{T}(\Omega)
\frac{\delta T}{T}(\Omega')
\Big\rangle
Y_{\ell m}^{\ast} (\Omega)
Y_{\ell' m'} (\Omega').
\label{queen}
\eea
The primordial perturbations
are related to
the observed CMB anisotropy by
\bea
\frac{\delta T}{T}(\Omega)
=\int
\frac{d^3k}{(2\pi)^3}
\sum_{L}\Big(\frac{2L+1}{4\pi}\Big)(-i)^L
 P_L ({\hat k}\cdot{\hat n})
\delta ({\bf k})\Theta_L (k),
\eea
where $\Theta_L(k)$ is the transfer function and
$\delta (\bf k)$ is the primordial perturbations in the Fourier space.
Here, we ignore the Integrated Sachs-Wolfe (ISW) effects
and focus on the Sachs-Wolfe (SW) one,
the transfer function is given by
\bea
\label{king2}
\Theta_\ell(k)\approx
-\frac{4\pi}{3}j_{\ell} (k \eta_\ast),
\eea
where $\eta_{\ast}:=\eta_0-\eta_{\rm ls}$,
$\eta_0$ and $\eta_{\rm ls}$ are the conformal times
of the present universe and of the photon last scattering, respectively.

The angular correlation function of the CMB fluctuations
is given by
\bea
\Big\langle
\frac{\delta T}{T}(\Omega)
\frac{\delta T}{T}(\Omega')
\Big\rangle
&=&
\int
\frac{d^3k}{(2\pi)^3}
\sum_{L,L'}
\frac{(2L +1)}{4\pi}
\frac{(2L'+1)}{4\pi}
(-i)^{L-L'}
P_L ({\hat k}\cdot{\hat n})
P_{L'}({\hat k}\cdot {\hat n}')
P(\bk)
\Theta_{L}(k)
\Theta_{L'}(k).
\label{acf}
\eea
Using the relation of the Legendre function
and the spherical harmonics
\bea
P_L ({\hat k}\cdot{\hat e})
=\frac{4\pi}{2L+1}
 \sum_{M=-L}^L
 Y_{LM}^{\ast}(\Omega)
 Y_{LM}(\Omega_k),
\eea
the angular correlation function
is given by
\bea
\Big\langle
\frac{\delta T}{T}(\Omega)
\frac{\delta T}{T}(\Omega')
\Big\rangle
&=&
\int
\frac{d^3k}{(2\pi)^3}
\sum_{L,L',M,M'}
(-i)^{L-L'}
Y_{LM}(\Omega)
 Y_{LM}^{\ast}(\Omega_k)
Y_{L'M'}^{\ast}(\Omega')
 Y_{L'M'}(\Omega_k)
P(\bk)
\Theta_L(k)
\Theta_{L'}(k).
\label{acf2}
\eea
Substituting \eqref{acf2} into \eqref{queen},
\bea
\langle
a_{\ell m}
a^{\ast}_{\ell' m'}
\rangle
&=&(-i)^{\ell-\ell'}
 \int
\frac{dk k^2}{(2\pi)^3}
\int d\Omega_k
 Y_{\ell m}^{\ast}(\Omega_k)
 Y_{\ell' m'}(\Omega_k)
P(\bk)
\Theta_{\ell}(k)
\Theta_{\ell'}(k),
\eea
where we have used the normalization condition Eq. (\ref{nc})
and $d^3k=k^2 dk d\Omega_k$.
Using the power spectrum \eqref{ansatz},
\bea \label{E12}
\langle
a_{\ell m}
a^{\ast}_{\ell' m'}
\rangle
&=&\delta_{\ell \ell'}
\delta_{m m'}
\frac{2}{\pi}
(4\pi)^{-2}
\int dk k^2 P_0(k)\Theta_\ell (k)^2
\nonumber\\
&+&
(-i)^{\ell-\ell'}
(4\pi)^{-2}
\frac{2}{\pi}
\int
dk k^2
 P_0(k)
\Theta_\ell (k)
\Theta_{\ell'}(k)
\Big(
\int d\Omega_k
 Y_{\ell m}^{\ast}(\Omega_k)
 Y_{\ell' m'}(\Omega_k)
 \g(k,\Omega_k)
\Big).
\eea

Since
the correction is negligible for the non-planar mode,
we may focus on the planar mode.
It is adequate to approximate the whole power spectrum
with that for the planar mode.
Since $\cos (2\Psi)$ term in the power spectrum Eq. (\ref{P:planar})
is oscillating,
we may ignore the angular power spectrum because of the rapid oscillation.
Thus the function $\g$ is given
\bea
\g(k,\Omega_k)
=\coth\Big(\frac{2^{2/3}\pi k}{3H_i}|\cos\theta_k|\Big)
-1.
\eea
Considering the SW approximation, Eq.~\eqref{E12} becomes
\bea
\langle
a_{\ell m}
a^{\ast}_{\ell' m'}
\rangle
&\approx&
\frac{2}{9\pi}
\Big\{
\delta_{\ell \ell'}
\delta_{m m'}
\int dk k^2 P_0(k)j_\ell (k \eta_\ast)^2
\nonumber\\
&+&(-i)^{\ell-\ell'}
\int dk k^2
 P_0(k)j_\ell (k \eta_\ast)j_{\ell'}(k \eta_\ast)
\int d\Omega_k
 Y_{\ell m}^{\ast}(\Omega_k)
 Y_{\ell' m'}(\Omega_k)
f (k,|\mu_k|)
\Big\} \,,
\label{step}
\eea
where $\mu_k=\cos\theta_k$ and
$
f (k,|\mu_k|)
:=\coth\Big(\frac{2^{2/3}\pi k}{3H_i}|\mu_k|\Big)
-1.$

Noting the definition of the spherical harmonic function
\bea
Y_{\ell m}(\Omega_k)
=(-1)^m\sqrt{\frac{2\ell+1}{4\pi}\frac{(\ell-m)!}{(\ell+m)!}}
  e^{im\phi_k} P_\ell^m(\cos\theta_k),
\eea
the integration along the azimuthal $\phi_k$-direction
reads
\bea
&&\int d\Omega_k
 Y_{\ell m}^{\ast}(\Omega_k)
 Y_{\ell' m'}(\Omega_k)
  f(k,|\mu_k|)
\nn\\
&&\quad=
\frac{\delta_{mm'}}{2}
\sqrt{(2\ell+1)(2\ell'+1)
\frac{(\ell-m)!(\ell'-m)!}
       {(\ell+m)!(\ell'+m)!} }
\int_{-1}^{1} d\mu_k
   P_\ell^m(\mu_k)
   P_{\ell'}^m(\mu_k)
 f(k, |\mu_k|).
\eea
We then perform the integration along the $\mu_k$-direction.
Noting the parity transformation of the Legendre functions
we can rewrite
\bea
\int_{-1}^{1} d\mu_k
   P_\ell^m(\mu_k)
   P_{\ell'}^m(\mu_k)
   f(k,|\mu_k|)
=\Big[
  1+ (-1)^{\ell+\ell'}
 \Big]
\int_{0}^{1} dr
   P_\ell^m(r)
   P_{\ell'}^m(r)
   f(k,r),
\eea
where $r:=\frac{k_3}{k}=|\mu_k|$.
Substituting  into \eqref{step}
\bea
\langle
a_{\ell m}
a^{\ast}_{\ell' m'}
\rangle
&\approx&
\frac{2\delta_{mm'}}{9\pi}
\Big\{
\delta_{\ell \ell'}
\int dk k^2 P_0(k)j_\ell (k \eta_\ast)^2
+
\frac{(-i)^{\ell-\ell'}}{2}\big[1+(-1)^{\ell+\ell'}\big]
\sqrt{(2\ell+1)(2\ell'+1)
 \frac{(\ell-m)!(\ell'-m)!}
        {(\ell+m)!(\ell'+m)!}}
\nonumber\\
&\times&
\int dk k^2
 P_0(k)j_\ell (k \eta_\ast)j_{\ell'}(k \eta_\ast)
\int_{r_{0}}^1 dr
  P_\ell^m (r)  P_{\ell'}^m (r)
  f(k,r)
\Big\},
\eea
where the lower bound on the $r$-integration
$0<r_0\ll 1$ is added as a cut-off
to avoid the divergence from $f(r_0\to 0,k)$.
The nonvanishing averaging is obtained for $m=m'$.

From now on, we
assume the exactly scale-invariant power spectrum
\bea
k^3P_0(k)= A_s,
\label{si}
\eea
where $A_s$ is
the amplitude of the primordial perturbations.
For later convenience,
we decompose the averaged multi-pole moments
into
the isotropic part and the anisotropic corrections
as
\bea
\langle
a_{\ell m}
a^{\ast}_{\ell' m'}
\rangle
=\delta_{mm'}
\Big\{
 C_{\ell}\delta_{\ell\ell'}
+\Delta C_{\ell \ell' m}(r_0)
\Big\},
\eea
where the isotropic part $C_\ell$ is given by
\bea
C_\ell=
\frac{2}{9\pi}
\int
dk k^2
P_0(k)
j_\ell(k\eta_\ast)^2,
\eea
and
\bea
\Delta C_{\ell \ell' m}(r_0)
&:=&\frac{(-i)^{\ell-\ell'}}{9\pi}
\big(1+(-1)^{\ell+\ell'}\big)
\sqrt{(2\ell+1)(2\ell'+1)
 \frac{(\ell-m)!(\ell'-m)!}
        {(\ell+m)!(\ell'+m)!}}
\nonumber\\
&\times&
\int_{k_{ir}}^{\infty} dk k^2
 P_0(k) j_\ell (k \eta_\ast)j_{\ell'}(k \eta_\ast)
\int_{r_{0}}^1 dr
  P_\ell^m (r)  P_{\ell'}^m (r)
  f(k,r),
\eea
respectively.
Here $k_{ir}$ is the IR cut-off for the anisotropic mode.
Using \eqref{si},
the isotropic part reduces to the standard result
\bea
\frac{\ell (\ell+1)}{2\pi}C_\ell
\approx \frac{A_s}{18\pi^2},
\label{sw_iso}
\eea
which gives the SW plateau seen for $\ell \lesssim 100$.

For a given $\ell$,
we take the ratio
\bea
\frac{\Delta C_{\ell \ell' m}(r_0)}{C_\ell}
&=&
(-i)^{\ell-\ell'}\ell(\ell+1)
\big(1+(-1)^{\ell+\ell'}\big)
\sqrt{
 \frac{(2\ell+1)(2\ell'+1)(\ell-m)!(\ell'-m)!}
        {(\ell+m)!(\ell'+m)!}}
\nonumber\\
&\times&
\int_{k_{ir}/H_i}^{\infty} \frac{dy}{y}
  j_\ell (y \eta_\ast H_i)j_{\ell'}(y \eta_\ast H_i)
\int_{r_{0}}^1 dr
  P_\ell^m (r)  P_{\ell'}^m (r)
\Big(\coth\Big(\frac{2^{2/3}\pi y r}{3}\Big)-1\Big),
\label{chi}
\eea
where 
we have changed the integration variable
$y=\frac{k}{H_i}$.
An observation
is that there is no contribution
from the anisotropic part
if  $\ell+\ell'$ is odd.
Thus we focus on the case $\ell+\ell'$ is even:

\begin{description}

\item{
1) $\ell={\rm even}$ and $\ell'={\rm even}$}

\item{
2) $\ell={\rm odd}$ and $\ell'={\rm odd}$.}

\end{description}

In the integrand in Eq. (\ref{chi}),
the dominant contribution comes from the region of $r\sim 0$.
In case of 1),
if $m={\rm even}$,
$P_\ell^{m}(r)\sim {\rm const}$ for a small $r$,
while
if $m={\rm odd}$,
$P_\ell^{m}(r)\sim r$
for a small $r$.
In case of 2),
if $m={\rm odd}$,
$P_\ell^{m}(r)\sim {\rm const}$
for a small $r$,
while
if $m={\rm even}$,
$P_\ell^{m}(r)\sim r$ for a small $r$.
Thus the integrand of $r$
is divergent as $r^{-1}$
in the cases
1) with an even $m$
and 2) with an odd $m$,
while it is proportional to $r$
in the cases 1) with an odd $m$
and 2) with an even $m$.
Thus we may focus on the first case,
to avoid too large anisotropic contribution
to the CMB angular spectrum.

We now focus on the case of $\ell'=\ell$.
From Eq. (\ref{chi}),
\bea
\label{eqa}
\frac{\Delta C_{\ell \ell m}(r_0)}{C_\ell}
&:=&
2\ell(\ell+1)
 \frac{(2\ell+1)(\ell-m)!}
        {(\ell+m)!}
\int_{k_{ir}}^{\infty} \frac{dk}{k}
  j_\ell (k \eta_\ast)^2
\int_{r_{0}}^1 dr
  P_\ell^m (r)^2
  \Big(\coth\Big(\frac{2^{2/3}\pi y r}{3}\Big)-1\Big).
\eea
Thus when $\ell+m$ is even,
the leading order contribution
from the $r$-integration is given by
\bea
&&\int_{k_{ir}/H_i}^{\infty}
\frac{dy}{y}
j_\ell  (H_i \eta_\ast y)^2
\int_{r_0}^1dr
P_{\ell}^m(r)^2
\Big(\coth\Big(\frac{2^{2/3}\pi y r}{3}\Big)-1\Big)
\nonumber\\
&\sim&
\frac{3\times 2^{2m}\ln r_0^{-1}}
       {2^{2/3}\Gamma\big(\frac{1}{2}(1-\ell-m)\big)^2
        \Gamma\big(1+\frac{1}{2}(\ell-m)\big)^2}
{\cal I} (
k_{ir} \eta_\ast),
\eea
where
\bea
{\cal I} (k_{ir} \eta_\ast)
&:=&\int_1^{\infty}
\frac{dy}{y^2}
    j_\ell (
k_{ir} \eta_\ast y)^2
=\frac{\pi}{(2\ell+3)(2\ell+1)(2\ell-1)
\Gamma\big(\frac{3}{2}+\ell \big)^2}
\Big\{
(
k_{ir} \eta_\ast)\Gamma\big(\frac{3}{2}+\ell \big)^2
\nonumber\\
&-&4^{-1-\ell}
(
k_{ir} \eta_\ast)^{2\ell}(3+8\ell+4\ell^2)
{}_2F_3
\Big[
\big\{
\ell-\frac{1}{2},
1+\ell
\big\},
\big\{
\ell+\frac{1}{2},
\ell+\frac{3}{2},
2\ell+2
\big\},
-(
k_{ir} \eta_\ast)^2
\Big]
\Big\}.
\label{length}
\eea
Note that
for $\ell \gtrsim H_i \eta_\ast$,
the second term in Eq. (\ref{length}) is negligible.
Hence from Eq. (\ref{eqa}), 
the leading order ratio becomes
\bea
\frac{\Delta C_{\ell\ell m}}{C_\ell}
&\sim&
{\tilde C} (\ell,m)
\left[(
k_{ir} \eta_\ast) \ln r_0^{-1}\right],
\eea
where
\bea
{\tilde C} (\ell,m)
&:=&
\frac{\ell(\ell+1)(\ell-m)!}
        {(\ell+m)!}
\frac{2^{1/3}\pi }
       {(2\ell+3)(2\ell-1)}
\frac{3\times 2^{2m} }
       {\Gamma\big(\frac{1}{2}(1-\ell-m)\big)^2
        \Gamma\big(1+\frac{1}{2}(\ell-m)\big)^2}.
\label{clm}
\eea
To avoid a too large anisotropic contribution,
$\Delta C_{\ell\ell m}/C_\ell$,
gives
a lower bound on $r_0$
\bea
r_0> {\rm exp}\Big[
-\frac{1}{{\tilde C}(\ell,m) (
k_{ir} \eta_\ast)}
\Big].
\eea

Before closing this Appendix,
we discuss the possible value of $ \eta_\ast$.
For simplicity, we ignore the contribution of the dark energy
and assume that the universe is matter-dominated
after the last scattering.
Representing the proper time in the matter-dominated universe
by $\tau_m$,
the scale factor
is given by
\bea
a_m(\tau_m)=a_0\Big(\frac{\tau_{m}}{\tau_{m,0}}\Big)^{\frac{2}{3}},
\eea
where $\tau_{m,0}$ is the present time and $a_0$ is the present size of the universe.
The conformal time, defined by  $d\eta=d\tau_m/a_m(\tau_m)$,
is now given by
\bea
\eta= \frac{3\tau_{m,0}^{2/3}\tau_{m}^{1/3}}{a_0}.
\eea
The comoving distance to the last scattering surface
is given by
\bea
\eta_\ast:=\eta_0-\eta_{\rm ls}
        =\frac{3\tau_{m,0}}{a_0}
\Big[1-\Big(\frac{\tau_{m,{\rm ls}}}{\tau_{m,0}}\Big)^{1/3}\Big]
       \simeq \frac{3\tau_{m,0}}{a_0},
\label{rast}
\eea
where $\frac{\tau_{m,{\rm ls}}}{\tau_{m,0}}\ll 1$.

The modes which contribute to the SW effects
reenter the horizon after the last scattering
$\tau_{m,{\rm ls}}<\tau_m<\tau_{m,0}$.
We assume that
the same modes
cross the horizon during inflation at the time $\tau$,
where the scale factor of the universe is given by
$a_{\rm inf}(\tau)=e^{H_i\tau}$,
and
the horizon exiting time during inflation $\tau$
and the re-entry time during the matter-dominated era
$\tau_m$ is related by
\bea
k= a_{\rm inf} (\tau) H_i= a_{m}(\tau_{m}) H_m (\tau_m),
\eea
where $H_m=a_m'(\tau_m)/a_m(\tau_m)=2/(3\tau_m)$.
Since
$a_m (\tau_m)H_m(\tau_m)=\frac{2a_0}{3\tau_{m,0}^{2/3} \tau_m^{1/3}}$,
we obtain
\bea
H_i= \frac{a_m H_m}{a_{\rm inf}}
  = \frac{2a_0}{3\tau_{m,0}^{2/3} \tau_{m}^{1/3}}e^{-H_i\tau},
\eea
which relates $\tau$ to $\tau_m$.
Assuming the mode which reenters the horizon today
exits the horizon at $\tau=\tau_0$ in the early universe,
by definition
\bea
H_i=  \frac{2a_0}{3\tau_{m,0}}e^{-H_i\tau_0}.
\eea
Using \eqref{rast}, we obtain
\bea
H_i \eta_\ast
\simeq
\Big(\frac{2a_0}{3\tau_{m,0}}e^{-H_i\tau_0} \Big)
\Big(\frac{3\tau_{m,0}}{a_0}\Big)
= 2e^{-H_i \tau_0}.
\eea
Note that the dependence on $a_0$
is canceled out.
If $\tau_{0}=0$, $H_i \eta_\ast=2$.
Assuming the largest mode we observe today crosses the horizon
just at the onset of inflation,
it may be natural to choose $\tau_0\sim 0$,
leading to $H_i \eta_\ast\sim 2$.
Thus the value of $H_i\eta_\ast$
should be taken around 2.


\end{document}